\documentclass[aps, twocolumn, prd, showkeys, preprintnumbers, superscriptaddress, nobibnotes, groupedaddress, noshowpacs, nofootinbib]{revtex4-1}

\usepackage{amsmath}
\usepackage{amsfonts}
\usepackage{amssymb}
\usepackage{bbold}
\usepackage{epsfig}
\usepackage{graphicx}
\usepackage{bm}
\usepackage{array}
\usepackage{listings}
\usepackage{color}
\usepackage[normalem]{ulem}
\definecolor{Blue}{RGB}{0, 0, 139}
\usepackage[colorlinks=true,allcolors=Blue,pdfusetitle]{hyperref}

\begin{document}

\title{Astrophysical constraints on non-standard coherent neutrino-nucleus scattering}

\author{Anna M. Suliga}
\email{anna.suliga@nbi.ku.dk}
\author{Irene Tamborra}
\email{tamborra@nbi.ku.dk}
\affiliation{
Niels Bohr International Academy and DARK, Niels Bohr Institute, \\
University of Copenhagen, Blegdamsvej 17, 2100, Copenhagen, Denmark
}

\date{\today}

\begin{abstract}
{
The exciting possibility of detecting supernova, solar, and atmospheric neutrinos with coherent neutrino-nucleus scattering detectors is within reach, opening up new avenues to probe New Physics. We explore the possibility of constraining non-standard coherent neutrino-nucleus scattering through astrophysical neutrinos. Sensitivity bounds on the mass and coupling of the new mediator are obtained by inspecting the modifications induced by the new interaction on the recoil rate observable in the upcoming RES-NOVA and DARWIN facilities. Under the assumption of optimal background tagging, the detection of neutrinos from a galactic supernova burst, or one-year exposure to solar and atmospheric neutrinos, will place the most stringent bounds for mediator couplings $g \gtrsim 10^{-5}$ and mediator masses between $1$ and $100$~MeV. A similar, but slightly improved, potential to COHERENT will be provided for larger mediator masses.
In particular, RES-NOVA and DARWIN may potentially provide one order of magnitude tighter constraints than XENON1T on the mediator coupling. Non-standard coherent neutrino-nucleus scattering may also force neutrinos to be trapped in the supernova core; this argument allows
to probe the region of the parameter space with $g \gtrsim 10^{-4}$, which is currently excluded by other coherent neutrino-nucleus scattering facilities or other astrophysical and terrestrial constraints. 
}
\end{abstract}

\maketitle

\section{Introduction}
\label{sec:Introduction}

Coherent Elastic Neutrino-Nucleus Scattering (CE$\nu$NS)~\cite{Freedman:1973yd} has recently been observed by the COHERENT Collaboration~\cite{Akimov:2017ade,Akimov:2018vzs}, despite the challenges due to the low recoil energy~\cite{Scholberg:2005qs}. Such a measurement is of paramount importance to test standard physics, but above all, it opens new avenues to probe physics beyond the Standard Model (SM)~{\cite{Dodd:1991ni,Scholberg:2005qs,Barranco:2005yy,Barranco:2007tz,Anderson:2012pn,Kosmas:2015sqa,Dutta:2015vwa,Dutta:2015nlo,Lindner:2016wff,Cerdeno:2016sfi,Lindner:2016wff,Dent:2016wcr,Liao:2017uzy,Coloma:2017ncl,Coloma:2017egw,Kosmas:2017tsq,Shoemaker:2017lzs,Dent:2017mpr,Farzan:2018gtr,Denton:2018xmq,AristizabalSierra:2019ykk,Khan:2019cvi,Cadeddu:2020nbr,Denton:2020hop,Tomalak:2020zfh,Flores:2020lji,Galindo-Uribarri:2020huw}.}

Coherent Elastic Neutrino-Nucleus Scattering is also at the core of direct detection dark matter experiments, such as XENON1T~\cite{Aprile:2015uzo}, LUX~\cite{Akerib:2016vxi}, XMASS~\cite{XMASS:2018bid} and the upcoming XENONnT~\cite{Aprile:2020vtw}, LZ~\cite{Akerib:2015cja}, and DARWIN~\cite{Aalbers:2016jon}. The advantage of CE$\nu$NS is the coherently enhanced cross section at low recoil energy coming from the square of the neutron number of the nucleus~\cite{Freedman:1977xn}. For this reason, the elastic scattering of neutrinos on protons or nuclei was proposed long ago as an attractive option to detect astrophysical neutrinos~\cite{Drukier:1983gj,Beacom:2002hs,Horowitz:2003cz}. 
This detection technique would be complementary to existing dedicated astrophysical neutrino experiments \cite{Scholberg:2012id,Nikrant:2017nya} and, being mediated through the Z boson, it would be flavor-insensitive.

Following up on this idea, Ref.~\cite{Lang:2016zhv} has highlighted the detection prospects of core-collapse supernova (SN) neutrinos in XENON1T and its future-generation extensions. In fact, typical SN neutrinos have an average energy of $\mathcal{O}(10)$~MeV, which would lead to a CE$\nu$NS nuclear recoil of $\mathcal{O}(1)$~keV, in the range of interest of  direct detection dark matter detectors based on Xenon (Xe)~\cite{Undagoitia:2015gya}. The transient nature of the SN burst and the high number of expected events would allow excellent detection perspectives, as also shown in Refs.~\cite{Raj:2019wpy,Raj:2019sci}. The detection of SN neutrinos through CE$\nu$NS has also been explored in LZ~\cite{Khaitan:2018wnf}, XMASS~\cite{XMASS:2016cmy}, and PICO-500~\cite{Kozynets:2018dfo}. 
In addition, the fact that direct detection dark matter experiments are approaching the so-called neutrino floor~\cite{Vergados:2008jp,Billard:2013qya,Boehm:2018sux}
could also be exploited to detect solar~\cite{Billard:2014yka,Cerdeno:2016sfi,Budnik:2017sbu,Newstead:2018muu} and atmospheric neutrinos~\cite{Newstead:2020fie}.

A new detector concept based on CE$\nu$NS has been recently presented to hunt for astrophysical neutrinos: RES-NOVA~\cite{Pattavina:2020cqc}. The employment of an array of cryogenic detectors based on archaeological lead (Pb) and the high Pb cross section with the ultra-high radiopurity of archaeological Pb promise high event statistics in RES-NOVA with easy scalability to large detector volumes.

Building on these new exciting developments, in this work we aim at placing constraints on the light mediator of CE$\nu$NS by adopting SN, solar, and atmospheric neutrinos. We exploit the effects that non-standard interactions could have both in the source and in the detector. In order to explore the largest region of the parameter space that could be eventually excluded, we focus on the perspective limits that could be set by the next-generation DARWIN~\cite{Aalbers:2016jon} and RES-NOVA-3~\cite{Pattavina:2020cqc}, but also discuss the bounds potentially provided by XENONnT~\cite{Aprile:2020vtw}, and XENON1T~\cite{Aprile:2019xxb}.

 This paper is organized as follows. In Sec.~\ref{sec:CENNS}, a brief overview on coherent neutrino nucleus scattering is provided, and the cross section is generalized to include non-standard vector and scalar mediators. Section~\ref{sec:Event rates} introduces the expected recoil rate in DARWIN and RES-NOVA. Section~\ref{sec:Supernova neutrinos} is centered on the impact of non-standard coherent neutrino nucleus scattering on the SN physics and on the detection of SN neutrinos. The effect of non-standard coherent neutrino nucleus scattering on the detection of solar and atmospheric neutrinos is shown in Secs.~\ref{sec:Solar neutrinos} and Sec.~\ref{sec:Atmospheric neutrinos}, respectively. The statistical analysis adopted to derive the bounds presented in this work is outlined in Sec.~\ref{sec:Statistical analysis}. A summary of our findings together with a discussion on  additional constraints on non-standard mediators in the light of  specific models is provided in Sec.~\ref{sec:Discussion}. Our conclusions and an outlook are presented in Sec.~\ref{sec:Conclusions}. We explore the impact of the uncertainty on the SN model on the sensitivity bounds in Appendix~\ref{Appendix_B}, and the dependence of the SN neutrino rate on the mediator mass in Appendix~\ref{Appendix_A}.

\section{Coherent Neutrino-Nucleus Scattering}
\label{sec:CENNS}

Coherent Neutrino-Nucleus Scattering occurs between an active neutrino flavor and a  nucleus
\begin{equation}
\label{eq:nucleus_scattering}
    \nu + T(A, Z) \rightarrow \nu + T(A, Z) \, 
\end{equation}
where $A, \ Z$ are the atomic mass and number of the nucleus $T$.
The differential CE$\nu$NS cross section in the Standard Model  is defined as~\cite{Freedman:1973yd}
\begin{equation}
\label{eq:sigma_nu_N}
    \frac{d\sigma_\mathrm{SM}}{dE_r} = \frac{G_F^2 m_T}{4\pi} Q_w^2 \left(1 - \frac{m_T E_r}{2 E_\nu^2} \right) F^2(Q) \ ,
\end{equation}
where $m_T = 931.5~\mathrm{MeV} \times A$ is the target nucleus $T$ mass, $Q_w =\left[N - Z(1 - 4\sin^2\theta_W)\right]$ is the weak vector nuclear charge, $\sin^2\theta_W = 0.231$~\cite{Patrignani:2016xqp} the Weinberg angle, $G_F$ is Fermi constant, $E_R$ is the recoil energy of the nucleus~$T$, $E_\nu$ the energy of a neutrino, and $Q = \sqrt{2 m_T E_r}$ is the momentum transfer. The maximum recoil energy is $E_r^\mathrm{max} = 2E_\nu^2 / (m_T + 2E_\nu)$. We employ the Helm-type form factor~\cite{Helm:1956zz,Kozynets:2018dfo}:
\begin{equation}
\label{eq:form_factor}
    F(Q) = 3 \frac{j_1(Q R_0)}{Q R_0} \exp{\left(-\frac{1}{2} Q^2 s^2\right)} \ ,
\end{equation}
with $j_1$ being the first order Bessel spherical function. The size of the nucleus $T$ is equal to $R_0 = \sqrt{R^2 - 5s^2}$ with $R = 1.2 A^{\frac{1}{3}}$, and the nuclear skin thickness is $s~\approx~0.5$~fm~\cite{Kozynets:2018dfo}. In what follows, we modify Eq.~\ref{eq:sigma_nu_N} in order to take into account non-standard vector and scalar mediators.

\subsection{New vector mediator}
\label{sec:New vector mediator}
The Lagrangian term for the new vector mediator ($Z^\prime$) coupling to neutrinos and quarks reads
\begin{equation}
\label{eq:Lagrangian_vector}
    \mathcal{L}^{Z^\prime} = g_{\nu,Z^\prime} Z^\prime_\mu \bar \nu_L \gamma^\mu \nu_L + Z^{\prime}_\mu \bar{q} \gamma^\mu g_{q, Z^\prime} q \ .
\end{equation}
In this case the cross section in Eq.~\ref{eq:sigma_nu_N} is modified as follows~\cite{Cerdeno:2016sfi,AristizabalSierra:2019ykk}
\begin{equation}
\label{eq:sigma_total_vector}
    \frac{d\sigma_{\nu N}}{dE_r} =  \frac{G_F^2 m_T}{\pi} |\xi|^2 \left( 1 - \frac{m_TE_r}{2E_\nu^2}\right) F^2(Q) \ ,
\end{equation}
where 
\begin{equation}
\label{eq:ksi}
   \xi = -\frac{Q_w}{2} + \frac{g_{\nu,Z^\prime} Q_w^\prime}{\sqrt{2}G_F (2m_T {E_r} + m_{Z^\prime}^2)} \ .
\end{equation}
In our framework we assume that the couplings to all quarks, $g_{q,Z^\prime}$, are such that $Q_w^\prime = g_{q, Z^\prime} 3 A$ \cite{Cerdeno:2016sfi}.

\subsection{New scalar mediator}
\label{sec:New scalar mediator}

For a scalar mediator ($\phi$), the new Lagrangian term is
\begin{equation}
\label{eq:Lagrangian_scalar_LNC}
    \mathcal{L}^{\phi} = g_{\nu,\phi} \phi \bar \nu_R \nu_L + \phi \bar q g_{q, \phi} q 
\end{equation}
for the lepton number conserving (LNC) coupling.
Analogously, the lepton number violating (LNV) Lagrangian term is 
\begin{equation}
\label{eq:Lagrangian_scalar_LNV}
    \mathcal{L}^{\phi} = g_{\nu,\phi} \phi \nu_L^c \nu_L + \phi \bar q g_{q, \phi} q \ .
\end{equation}
Note that, due to the chirality-flipping nature of the new interaction  (for both LNC and LNV couplings), the scattering may be no longer elastic as the outgoing neutrino may have opposite handness, which means that a sterile neutrino may be produced in the LNC case 
and antineutrino in the LNV case. Hence, in the following, we will use C$\nu$NS (instead than CE$\nu$NS) when dealing with the scalar mediator. 

The total cross section in the presence of the new scalar mediator is~\cite{AristizabalSierra:2019ykk,Farzan:2018gtr}:
\begin{equation}
\label{eq:sigma_scalar}
    \frac{d\sigma_{\nu N}}{dE_r} = \frac{d\sigma_\mathrm{SM}}{dE_r} + \frac{d\sigma_{\phi}}{dE_r} \ ,
\end{equation}
where $d \sigma_\mathrm{SM}(E_\nu) / dE_r$ is defined as in Eq.~\ref{eq:sigma_nu_N} and the  term related to the new mediator is~\cite{AristizabalSierra:2019ykk,Farzan:2018gtr} 
\begin{equation}
\label{eq:sigma_total_scalar}
    \frac{d \sigma_\mathrm{\phi}}{dE_r} = \frac{(g_{\nu, \phi} g_{q, \phi} Q_s)^2 }{2 \pi (2 E_r m_T + m_\phi^2)^2} \; \frac{m_T^2 E_r}{2E_\nu^2} \; F^2(Q) \ ,
\end{equation}
with $Q_s =\sum_{N,q} m_N/m_q f^{N}_{T_q} \approx 14A + 1.1Z$;  the hadronic form factors $f^{N}_{T_q}$ follow the ones used in Ref.~\cite{Cerdeno:2016sfi}.  
Throughout the paper, if not otherwise specified, we always assume  $g = \sqrt{|g_{q,i} g_{\nu, i}|}$, where $g_{q,i} g_{\nu, i} > 0$ and $i = \{Z^\prime, \phi\}$.

\section{Event rates in RES-NOVA and DARWIN}
\label{sec:Event rates}
The interaction of astrophysical neutrinos through CE$\nu$NS (and C$\nu$NS for a scalar mediator, see Sec.~\ref{sec:New scalar mediator}) in the detector, in the presence of non-standard interactions, is described by the following differential  rate:
\begin{equation}
\label{eq:Event_rate_nu_N}
\begin{aligned}
    \frac{dR_{\nu N}}{dE_r dt} &= N_T \; \epsilon(E_r) \\
    & \times  \int dE_{\nu} \ \frac{d\sigma_{\nu N}}{dE_r} \ \psi(E_\nu, t) \ \Theta (E_r^\mathrm{max} - E_r) \ ,
\end{aligned}
\end{equation}
where $N_T$ is the number of target nuclei in the detector volume, $\epsilon(E_r)$ is the detector efficiency, $\psi(E_\nu)$ is the neutrino flux,  $\Theta$ is the Heaviside step function, and $E_r^\mathrm{max}$ is defined as in Sec.~\ref{sec:CENNS}.

In this work, we mainly focus on RES-NOVA and DARWIN, which are the largest upcoming detectors employing CE$\nu$NS. 
In order to estimate the effect of non-standard CE$\nu$NS (and C$\nu$NS) occurring in the detector,  we explore the effects induced on the recoil rate by non-standard interactions of astrophysical neutrinos in Pb and Xe detector in Secs.~\ref{sec:Supernova neutrinos}--\ref{sec:Atmospheric neutrinos} by relying on  the scattering rates per ton-year per target material. We will then  apply our findings to RES-NOVA and DARWIN in Sec.~\ref{sec:Statistical analysis}. In the following, we introduce the general features of these detectors.

\subsection{RES-NOVA}

RES-NOVA is a {recently proposed  detector that, if funded, might be installed} at the Gran Sasso National Laboratories~\cite{Pattavina:2020cqc}. The sensitive detector component should be made of archaeological Pb.
The ultra-low radioactivity of archaeological Pb and  the largest CE$\nu$NS cross section of Pb, among the stable elements, have the potential to guarantee high statistics. 

Unless otherwise specified, we assume a detector effective volume of $456$~ton (corresponding to RES-NOVA-3, planned after the first two extensions: RES-NOVA-1,2), $100\%$ detection efficiency, and $1$~keV recoil energy threshold. We refer the reader to Ref.~\cite{Pattavina:2020cqc} for more details.

In what follows, we assume negligible backgrounds in RES-NOVA. The irreducible detector background, originating from the decay of radioactive isotopes, is neglected; this assumption is justified for the detection of  SN neutrinos, since the background rate of archeological Pb is several orders of magnitude lower than the SN signal rate. In the case of atmospheric and solar neutrinos, this background can be reduced significantly by relying on particle identification techniques~\cite{Pattavina:2020cqc}. In addition, an efficient suppression of the cosmogenic neutron background can also be achieved~\cite{Pattavina:2020cqc}.

\subsection{DARWIN}
Dual-phase Xe detectors are sensitive to sub-keV nuclear recoil energies, with very low background and excellent time resolution.
The detection of the prompt scintillation light (the S1 signal) from the liquid phase of the detector and the delayed scintillation photons (the S2 signal), caused by the drifting ionization electrons in the gaseous phase of the detector, allows for disentangling the electronic and nuclear recoils based on the S2/S1 ratio. 
DARWIN is expected to succeed XENON1T, and XENONnT at Gran Sasso National Laboratories.

DARWIN will have an effective volume of $40$~ton. The detector efficiency is estimated to be between 100$\%$ and the energy-dependent one shown in Fig.~1 of Ref.~\cite{Aprile:2018dbl,Aprile:2019xxb}; 
the energy threshold is of $1$~keV~\cite{Aalbers:2016jon} (see Refs.~\cite{Aprile:2018dbl,Aalbers:2016jon,Aprile:2019xxb} for more details).

Similarly to RES-NOVA, the irreducible detector backgrounds for neutrino detection in DARWIN are arising from cosmogenic neutrons and the radioactivity of the intrinsic detector materials.
Again, these backgrounds can be safely neglected for the study of SN neutrinos~\cite{Lang:2016zhv}. For the detection of solar and atmospheric neutrinos, the cosmogenic background in DARWIN can be eliminated to a very good extent by means of an active veto, self-shielding, and by rejection of multiple scattering 
events~\cite{Aalbers:2016jon}. The low radioactivity of the detector material will be achieved with purification techniques and rejection techniques based on looking at the S2 to S1 signal 
ratio~\cite{Aalbers:2016jon}. 

Ours is an exploratory study where we aim at comparing the sensitivity to New Physics of RES-NOVA and DARWIN {for generic non-standard coherent scattering scenarios}. Hence, we refrain from estimating the S1 and S2 signals detectable in DARWIN and focus on the nuclear recoil event rates. It should also be noted that the distinction between the nuclear and electronic recoils has already been achieved in XENON1T~\cite{Aprile:2018dbl,Aprile:2019xxb} and its predecessors.

\section{Supernova neutrinos}
\label{sec:Supernova neutrinos}

Core-collapse SNe are among the densest neutrino sources~\cite{Vitagliano:2019yzm,Mirizzi:2015eza}. In this section, we first place constraints on non-standard neutrino nucleon coherent scattering in the source and then focus on the non-standard interactions of SN neutrinos occurring in the detector.

{Since our main goal is to investigate the sensitivity of C$\nu$NS to New Physics,  
 we focus on the modifications induced by the non-standard $\nu + N \rightarrow \nu +N$ reaction on the SN physics. 
 This interaction channel allows us to place constraints on $g=\sqrt{g_\nu g_q}$. We neglect the contribution from the non-standard neutrino Bremsstrahlung, as it is expected to be a sub-leading process.

The $\nu + N \rightarrow \nu +N$ interaction channel also enables us to directly compare the SN limits to the ones obtained from C$\nu$NS detectors (see Sec.~\ref{sec:Statistical analysis}). 
Additional SN constraints, stemming from interactions involving only neutrino or quark couplings, might be relevant depending on the details of the considered model (see Sec.~\ref{sec:Discussion} for a discussion). 
}

\subsection{Non-standard coherent scattering in the supernova core}
\label{sec:Effects on the neutrino diffusion time}

\begin{figure*}[t]
    \centering
    \includegraphics[scale=0.45]{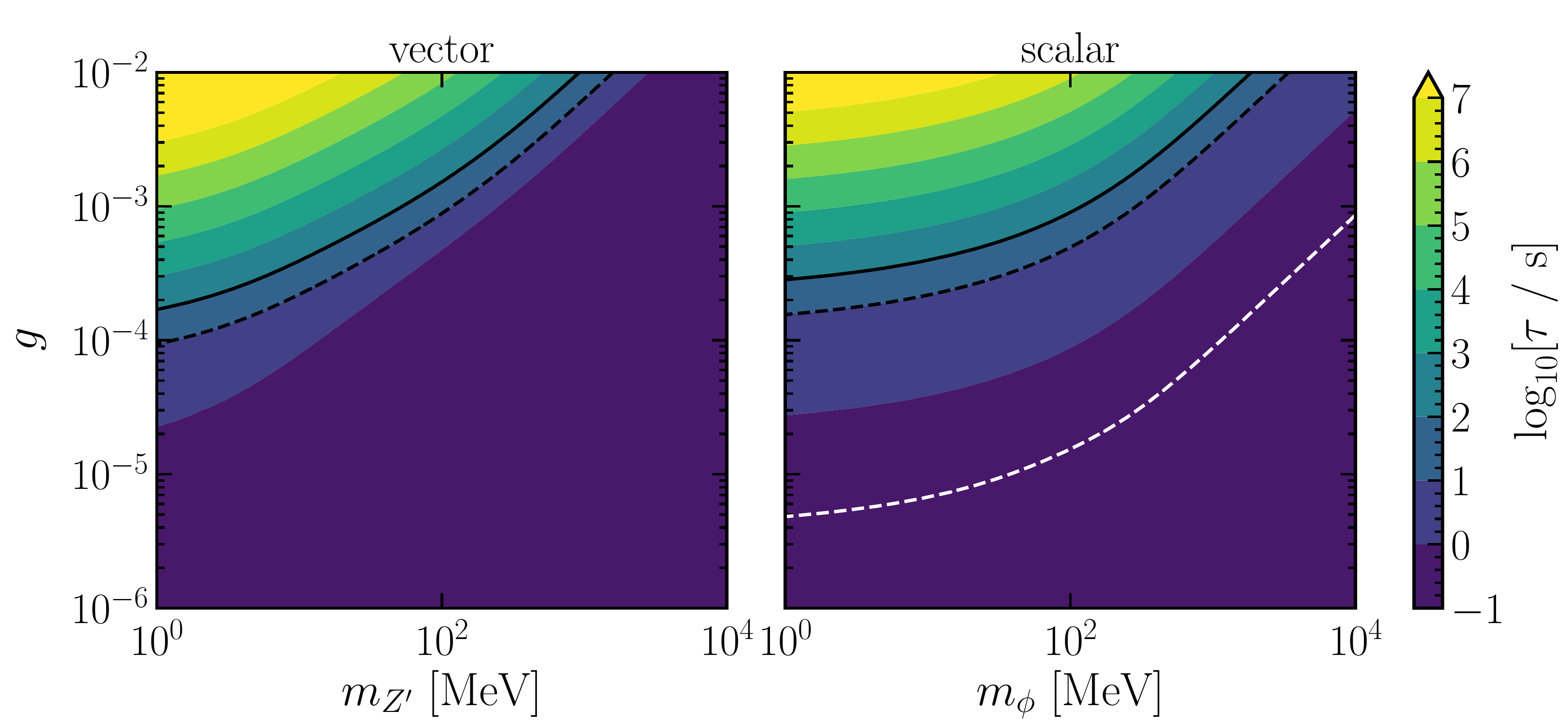}
    \caption{Contour plot of the average diffusion time of neutrinos in the SN core in the ($g, m_{Z^\prime}$ [$g, m_{\phi}$]) parameter space for the vector [scalar] mediator on the left [right]. The black lines mark the average diffusion time $ \tau = 10$~s (dashed) and $100$~s (solid);  see text for  details. For the LNC scalar mediator we also show the white (grey) line above which $\nu_e$ ($\nu_x$) are expected to scatter on nucleons at least once (cooling argument, see text). The regions of the parameter space above the black and white curves are excluded.}
    \label{fig:2D_SN_INSIDE_2}
\end{figure*}

In the SN core, the matter density is large enough [$\rho \simeq \mathcal{O}(10^{14})$~g~cm$^{-3}$] that neutrinos are trapped. If neutrinos undergo non-standard coherent scattering with the nucleons, for large enough couplings, they could be trapped for too long in the SN core, and hence fail in reviving the shock, according to the delayed neutrino-driven explosion mechanism~\cite{Bethe:1984ux}. 
In order to estimate the diffusion time of neutrinos, first we calculate the mean distance that a neutrino can travel between  interactions, i.e., the average mean-free path:
\begin{equation}
\label{eq:mean_free_path}
\begin{aligned}
    \lambda_{\nu_\beta} = \sum_{\mathrm{CC, NC}} \frac{\int dE_{\nu_\beta} f(E_{\nu_\beta}) E_{\nu_\beta}^2}{n_t \int dE_{\nu_\beta} f(E_{\nu_\beta}) E_{\nu_\beta}^2 \sigma_i(E_{\nu_\beta})} \ ,
\end{aligned}
\end{equation}
where $f(E_{\nu_\beta})$ is the Fermi-Dirac distribution for the $\beta$-flavor (anti)neutrino, $n_t$ is the number of targets for  charged-current (CC) or neutral current (NC) interactions. Given that we focus on nuclear densities, we assume that the scattering on nucleons is the driving contribution in determining the mean-free path ($\nu_e + n \rightarrow e^- + p$, $\bar\nu_e + p \rightarrow e^+ + n$, and $\nu_\beta + N \rightarrow \nu_\beta + N$). The CC cross sections have been implemented following Ref.~\cite{Strumia:2003zx}. The standard and non-standard NC scattering on neutrons  [protons] have been estimated by replacing $Q_w \rightarrow 1+3g_A$ [$(1-4\sin^2\theta_W) + 3g_A$]~\cite{Giunti:2007ry} in Eqs.~\ref{eq:sigma_nu_N}, \ref{eq:sigma_total_scalar} and \ref{eq:sigma_total_vector} and assuming $F(Q^2) \approx 1$.

The diffusion time of neutrinos is 
\begin{equation}
\label{eq:diffusion_time}
    \tau_{\nu_\beta} = \int_{R_1}^{R_2} dr \frac{r}{\lambda_{\nu_\beta}(r)} \ , 
\end{equation}
where $R_1$ defines the inner boundary where  neutrino degeneracy becomes negligible and Pauli blocking is not large anymore, and $R_2$ stands for the outer radius where neutrinos start to free stream. In the following, we focus on  electron neutrinos and antineutrinos as they are mainly responsible for depositing energy to revive the stalled shock~\cite{Mirizzi:2015eza}.
To estimate the diffusion time, we rely on a time snapshot ($t_\mathrm{pb} = 0.25$~s) from a one-dimensional spherically symmetric SN simulation with  18.6~$M_\odot$ mass~\cite{Garc:SN} and SFHo \cite{Steiner:2012rk} nuclear equation of state. Note that these inputs should be considered as indicative in order to gauge the impact of non-standard coherent scattering and do not qualitatively affect our findings.

Figure~\ref{fig:2D_SN_INSIDE_2} shows a contour plot of the average diffusion time of (anti)neutrinos in the plane spanned by $m$ and $g$, for the vector (scalar) mediator on the left (right). The average diffusion time is defined as $\tau = 1/2 (\tau_{\nu_e} + \tau_{\bar \nu_e})$, and it is the time required for electron neutrinos and antineutrinos to travel between $R_1 = 10$~km and $R_2 = 40$~km. We have chosen $R_1$ such that the degeneracy parameter $\eta = \mu / T =1$ (with $\mu$ and $T$ being the $\nu_e$ chemical potential and medium temperature, respectively) and hence Pauli blocking is negligible; $R_2$  coincides with the neutrinosphere (i.e.,~the surface above which neutrinos free stream). In the presence of non-standard coherent scattering, we exclude the region of the parameter space such that  $\tau \gtrsim 10~(100)$~s. In this circumstance, neutrinos will be trapped for too long, possibly halting the shock revival, and this would be in conflict with the neutrino signal observed from  SN 1987A~\cite{Hirata:1987hu,Bionta:1987qt}.

The region of the parameter space where the new mediator coupling highly affects the diffusion time of neutrinos, making it longer than $\tau = 10~(100)$~s, lays above the black dashed (solid) lines in Fig.~\ref{fig:2D_SN_INSIDE_2}. The excluded region is smaller for the  vector mediator case because of the presence of the interference term in the cross section (Eq.~\ref{eq:ksi}) that reduces the NC contribution to the diffusion time when the standard coupling is comparable to the non-standard one.

The  scalar mediator case is more complex. In fact, the diffusion time criterion is applicable both for LNC and LNV couplings. However, for LNV coupling, neutrinos can convert to antineutrinos. As a result, the new interactions can have a drastic impact on the proto-neutron star evolution~\cite{Farzan:2018gtr} and, possibly, on neutrino flavor conversions eventually occurring before neutrino decoupling~\cite{Chakraborty:2016yeg}.  However, a self-consistent modeling of these effects is out of reach at the moment and it is beyond the scope of this paper.

For the LNC scalar mediator, left-handed neutrinos convert to right-handed ones (effectively becoming steriles). In this case,  the total number of neutrino-nucleon scatterings according to the central limit theorem~\cite{Chandrasekhar:1943ws} is
\begin{equation}
\label{eq:number_of_scatters}
N = \int_0^{R_2}  \frac{2 r }{\lambda(r)^2} \; dr  \ .
\end{equation}
If all neutrinos, on average, scatter once before reaching the neutrinosphere ($R_2$), it means that they are converted to  right-handed particles. As such, there should be an extra cooling channel~\cite{Raffelt:1996wa} that would shorten the timescale of the neutrino signal otherwise observed  from the SN 1987A or even cause it to disappear. 
Hence, in the right panel of Fig.~\ref{fig:2D_SN_INSIDE_2}, we exclude the region of the parameter space such that $N \gtrsim 1$. This argument excludes a bigger fraction of the ($g$, $m_\phi$) parameter space than the diffusion argument.

Despite the fact that we rely on inputs from one hydrodynamical SN simulation and develop detailed calculations of the neutrino mean-free path, diffusion time, and number of neutrino-nucleon interactions, our findings are in good agreement with the ones of Ref.~\cite{Farzan:2018gtr}. In Ref.~\cite{Farzan:2018gtr} the authors estimated the bounds on the mediator mass and coupling for a constant matter density and a single neutrino energy, using our same arguments.

\subsection{Non-standard coherent scattering of supernova neutrinos in the detector}
\label{sec:SN Effects on the event rates}
\begin{figure*}[t]
    \centering
    \includegraphics[scale=0.45]{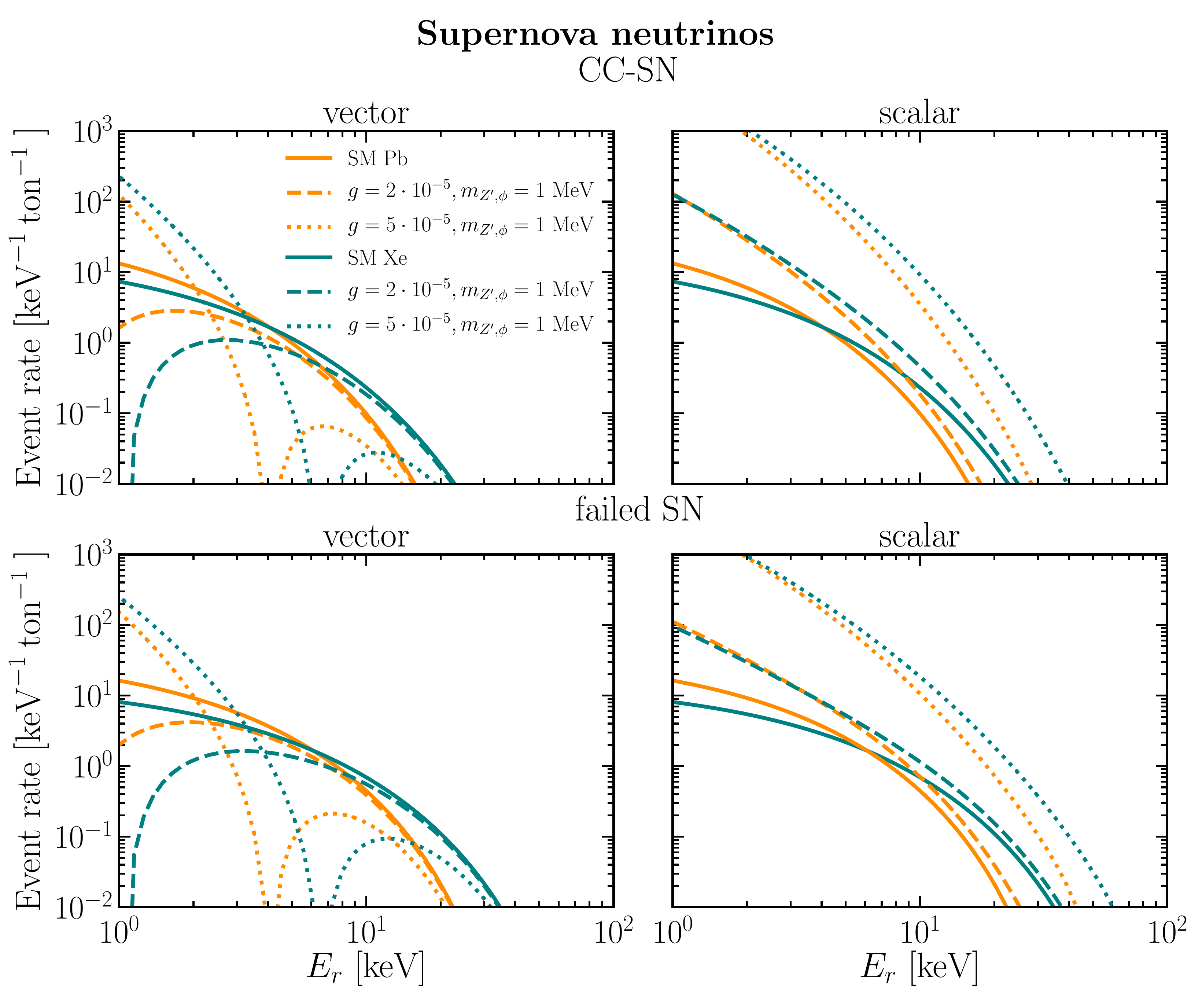}
    \caption{Expected event rate as a function of the nuclear recoil energy for a CC-SN (top panels) and a failed SN (bottom panels) at $10$~kpc from Earth in Pb (orange curves) and Xe (blue curves) based detectors. The standard event rates (solid lines) are shown together with  the non-standard rates for vector and scalar mediators, on the left and right panels, respectively. The scalar mediator only increases the event rate with respect to the standard case; this is not true for the vector mediator case.}
    \label{fig:Rate_SN_BMS}
\end{figure*}

In order to estimate the event rate of SN neutrinos expected in RES-NOVA and DARWIN, we rely on the outputs of two one-dimensional spherically symmetric core-collapse SN models~\cite{Garc:SN}. Since  core-collapse SNe may be as likely as black hole forming collapses~\cite{Sukhbold:2015wba,Basinger:2020iir}, we adopt  a core-collapse SN model with $27~M_\odot$ (CC-SN) and a $40~M_\odot$ black-hole-forming collapse model (failed SN) as inputs for our computations. 
The neutrinos emitted during the SN burst have a pinched thermal energy distribution~\cite{Keil:2002in,Tamborra:2012ac}
\begin{equation}
\label{eq:Sn_spectrum}
\begin{aligned}
\phi_{\nu_\beta} (E_{\nu_\beta}, t_{\mathrm{pb}}) =& \;\xi_{\nu_\beta} (t_\mathrm{pb}) \left( \frac{E_{\nu_\beta}}{\langle E_{\nu_\beta} (t_\mathrm{pb}) \rangle} \right)^{\alpha_\beta (t_\mathrm{pb})} \\
& \times \exp \left[-\frac{(1+\alpha_\beta (t_\mathrm{pb})) E_{\nu_\beta}}{\langle E_{\nu_{\beta}}(t_\mathrm{pb}) \rangle} \right] \ ,
\end{aligned}
\end{equation}
where the pinching factor is given by 
\begin{equation}
\label{eq:pinching_factor}
\alpha_\beta (t_\mathrm{pb}) = \frac{2 \langle E_{\nu_\beta} (t_\mathrm{pb}) \rangle ^2 - \langle E_{\nu_\beta} (t_\mathrm{pb})^2 \rangle}{\langle E_{\nu_\beta} (t_\mathrm{pb})^2 \rangle - \langle E_{\nu_\beta} (t_\mathrm{pb}) \rangle^2}  \ .  
\end{equation}
The time-integrated neutrino flux is calculated as
\begin{equation}
\label{eq:Sn_flux}
    \psi (E_{\nu_\beta}) = \int_{0}^{t_\mathrm{max}} dt_\mathrm{pb} \; \frac{L(t_\mathrm{pb})}{4 \pi D^2} \frac{\phi_{\nu_\beta} (E_{\nu_\beta}, t_\mathrm{pb})}{\langle E_{\nu_\beta(t_\mathrm{pb})} \rangle} \ ,
\end{equation}
where $D$ is the SN distance from Earth; in the following, we consider $t_\mathrm{max}= 14$~s and $t_\mathrm{max} = 0.2$~s for the CC-SN and failed SN respectively, and $D=10$~kpc unless otherwise stated. Since the CE$\nu$NS (C$\nu$NS) interaction is not flavor sensitive, we do not worry about neutrino flavor conversions and consider the total neutrino flux summed over six flavors.

Figure~\ref{fig:Rate_SN_BMS} illustrates the event rate of SN neutrinos for Pb based (orange color) and Xe based (green lines) detectors for standard (solid lines) and non-standard interactions (non-solid lines) for the CC-SN [failed SN] on the top [bottom] panels. 
The number of events of the black-hole-forming collapse at high recoil energies (above the detection threshold) exceeds the number of events from the core-collapse SN. This is because of the hotter neutrino spectra emitted in   black-hole-forming collapses~\cite{Mirizzi:2015eza}. The Pb based detector is characterized by a total number of events above threshold ($1$~keV) higher than the Xe one; this is due to the fact that the event rate scales as the number of neutrons of the target.
However, due to the heavier target, the range of observed recoil energies is smaller for Pb.

The left (right) panels of Fig.~\ref{fig:Rate_SN_BMS} illustrate the  modifications in the expected event rate for different values of $g$ and $m_{Z^\prime,\phi}=1$~MeV. 
The  number of events as a function of the recoil energy for the new scalar interaction increases with respect to the standard case. In contrast, the event rate can be smaller than in the standard case for the vector mediator scenario. This can be easily understood by comparing the total cross sections for  scalar (Eq.~\ref{eq:sigma_total_scalar}) and vector (Eq.~\ref{eq:sigma_total_vector}) mediators. In the fist case, the cross section can only increase, if the quark and neutrino couplings to the new mediator are positive. Yet, for the vector mediator, the total cross section is the sum of the contributions from the new vector mediator and the $Z$ boson; hence, the interference term between these two can lead to a depletion of the event rate when the interference term dominates.  
\begin{figure*}[t]
    \centering
    \includegraphics[scale=0.45]{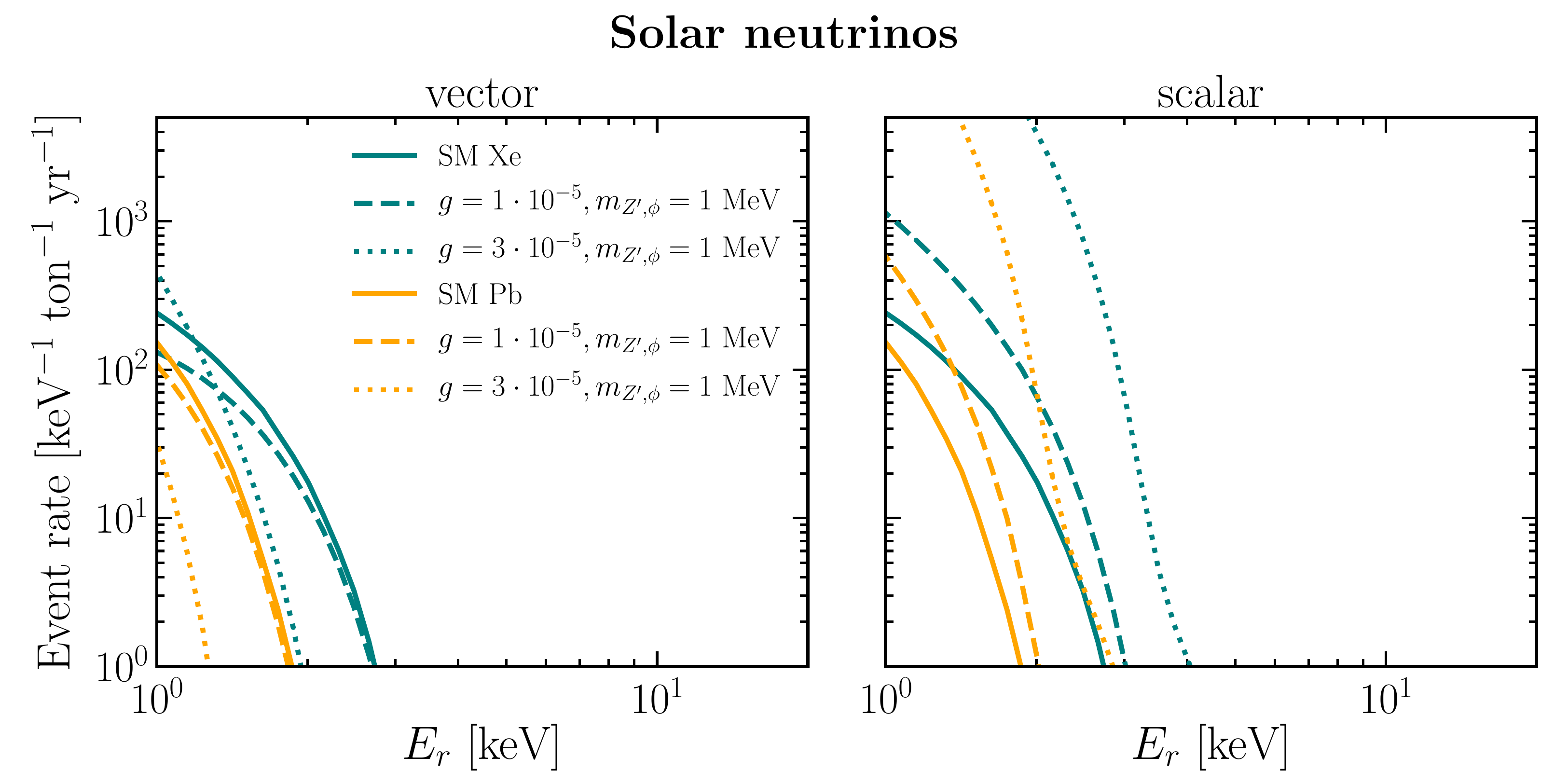}
    \caption{Expected event rate as a function of the nuclear recoil energy induced by  solar neutrinos in Pb (orange) and Xe (blue) based detectors. The standard event rate (solid line) is shown together with the non-standard one  for  the vector and scalar mediator on the left and right panels, respectively. The Xe based detector has an advantage over the Pb based detector thanks to the recoil spectrum extending to higher energies.}
    \label{fig:Solar_BSM}
\end{figure*}

\section{Solar neutrinos}
\label{sec:Solar neutrinos}

The sun is the closest astrophysical source of neutrinos, see Ref.~\cite{Vitagliano:2019yzm} for a recent overview and references therein. 
Solar neutrinos are very useful to constrain non-standard scenarios, see e.g.~Refs.~\cite{Vergados:2008jp,Harnik:2012ni,Cerdeno:2016sfi,Budnik:2017sbu,Dutta:2017nht,AristizabalSierra:2017joc,Boehm:2018sux,Gonzalez-Garcia:2018dep}. 
In principle, bounds on light mediators coupling to neutrinos and quarks may be derived by looking at neutrino interactions with the nucleons inside the sun. However, the solar density [$\rho \simeq \mathcal{O}(10$--$100)~\mathrm{g}/\mathrm{cm}^{-3}]$ is too low to trap neutrinos even for large $g$, hence the non-standard coherent neutrino-nucleus scattering affects the physics of the sun negligibly, differently from what discussed in the SN case in Sec.~\ref{sec:Effects on the neutrino diffusion time}.

The detection of solar neutrinos in Pb and Xe based detectors
could, however, allow one to constrain non-standard coherent neutrino nucleus scattering. This option is especially interesting since, as we approach the neutrino floor in direct detection dark matter experiments, the detection of solar neutrinos is within reach~\cite{Vergados:2008jp,Cerdeno:2016sfi}. 
In order to estimate the event rate of solar neutrinos, we consider the flux of neutrinos coming from two nuclear processes in the Sun 
\begin{equation}
    {}^{8}\mathrm{B} \rightarrow {}^{8}\mathrm{Be}^* + e^+ + \nu_e
\end{equation}
and
\begin{equation}
    {}^{3}\mathrm{He} + p \rightarrow {}^{4}\mathrm{He} + e^+ + \nu_e \ .
\end{equation}
These reactions produce neutrinos with energy up to 15~MeV. The rest of the neutrinos made in the $pp$ and CNO chains, that fuel the sun, create neutrinos with typical energies that are too low to produce nuclear recoils above the $1$~keV threshold. The solar neutrino fluxes adopted in this work were obtained by averaging the lower and upper bounds provided in the tables of Ref.~\cite{Vitagliano:2019yzm}.

The expected rate of solar neutrinos is shown in Fig.~\ref{fig:Solar_BSM} for Xe and Pb targets  for the standard case (solid lines). Similarly to the SN event rate (Fig.~\ref{fig:Rate_SN_BMS}), the recoil spectrum ends at a smaller recoil energy for the Pb detector because the maximum recoil energy ($E_r^\mathrm{max}$) is smaller for a heavier detector material. Yet, due to the fact that the maximum neutrino energy produced by the nuclear reactions in the sun ($\sim15$~MeV) is much smaller than in the SN case ($\sim 50$~MeV), we can see that the expected recoil energies produced by solar neutrinos lay below $5$~keV. For the same reason, the crossing between the event rates of the Xe and Pb based detectors lays below the $1$~keV threshold, causing the Xe rate to be higher than the Pb one in the plotted energy range.

The impact of a non-standard mediator with $1$~MeV mass in the interactions of solar neutrinos with the detector targets is displayed in Fig.~\ref{fig:Solar_BSM} for different couplings $g$, on the left for the vector mediator and on the right for the scalar one. In both cases, the event rate is shown for Xe and Pb targets.
The maximum recoil energy in the case of the Xe based detector is almost twice  the one of the Pb detector;
this allows  to look for energy-dependent features in the recoil spectra over a broader range of energies in Xe based detectors. As we will discuss in Sec.~\ref{sec:Statistical analysis}, this opportunity helps to constrain the vector mediator at low and intermediate masses (see Fig.~\ref{fig:2D_SUN_ATM}). This is not as important for the scalar mediator because in the low mediator mass limit the non-standard cross section (Eq.~\ref{eq:sigma_scalar}) does not depend on the mediator mass nor on the target mass. One can also notice that the event rate for the scalar mediator is always increasing with respect to the standard one, contrarily to the vector case, as dictated by the cross sections in Eqs.~\ref{eq:sigma_total_vector} and \ref{eq:sigma_scalar}. 
In addition, for the same coupling, the difference between the standard and non-standard event rate is bigger for the scalar mediator in the low mass limit, due to the lack of neutrinos with energies higher than $\sim 15$~MeV and the $E_r^{-1} E_\nu^{-2}/2$ dependence  of the scalar cross section as opposed to $(1-m_T E_r E_\nu^{-2}/2)$ in the interference term of vector cross section.

\section{Atmospheric neutrinos}
\label{sec:Atmospheric neutrinos}

Atmospheric neutrinos originate from the decay of mesons produced by the interactions of cosmic rays with nuclei in the Earth's atmosphere. The atmospheric neutrino flux at sub-GeV energies mainly consists of neutrinos from  charged pion decays~\cite{Newstead:2020fie}. In the decay of a positively charged pion, a muon (anti)neutrino is created together with an electron neutrino:
\begin{equation}
\begin{aligned}
    \pi^{+} \rightarrow \mu^{+}  + \nu_\mu \ \mathrm{and}\ 
    \mu^{+} \rightarrow  e^{+} + \bar \nu_\mu + \nu_e \ .
\end{aligned}
\end{equation}
Similarly, for the negatively charged pion, two antineutrinos $\bar \nu_e$ and $\bar \nu_\mu$ and an electron neutrino are formed.   
 {Theoretical predictions on low energy atmospheric neutrinos are still plagued by high uncertainty ($\sim 20\%$)~\cite{Honda:2011nf,Agrawal:1995gk}; the latter, for neutrinos with energy below $100$~MeV, might increase up to $50\%$~\cite{Battistoni:2002ew}. }

Future generation CE$\nu$NS detectors  will be able to observe the atmospheric neutrino background~\cite{Newstead:2020fie,Pattavina:2020cqc}. We capitalize on this opportunity to explore how the presence of a non-standard coherent  mediator could change the expected event rates.
We rely on the inputs for the atmospheric neutrino flux provided in Ref.~\cite{Newstead:2020fie}, where the atmospheric neutrino flux was calculated {by using the FLUKA code~\cite{Battistoni:2002ew}} for the Gran Sasso location, where RES-NOVA and DARWIN {may be built, if approved}.

\begin{figure*}[t]
    \centering
    \includegraphics[scale=0.45]{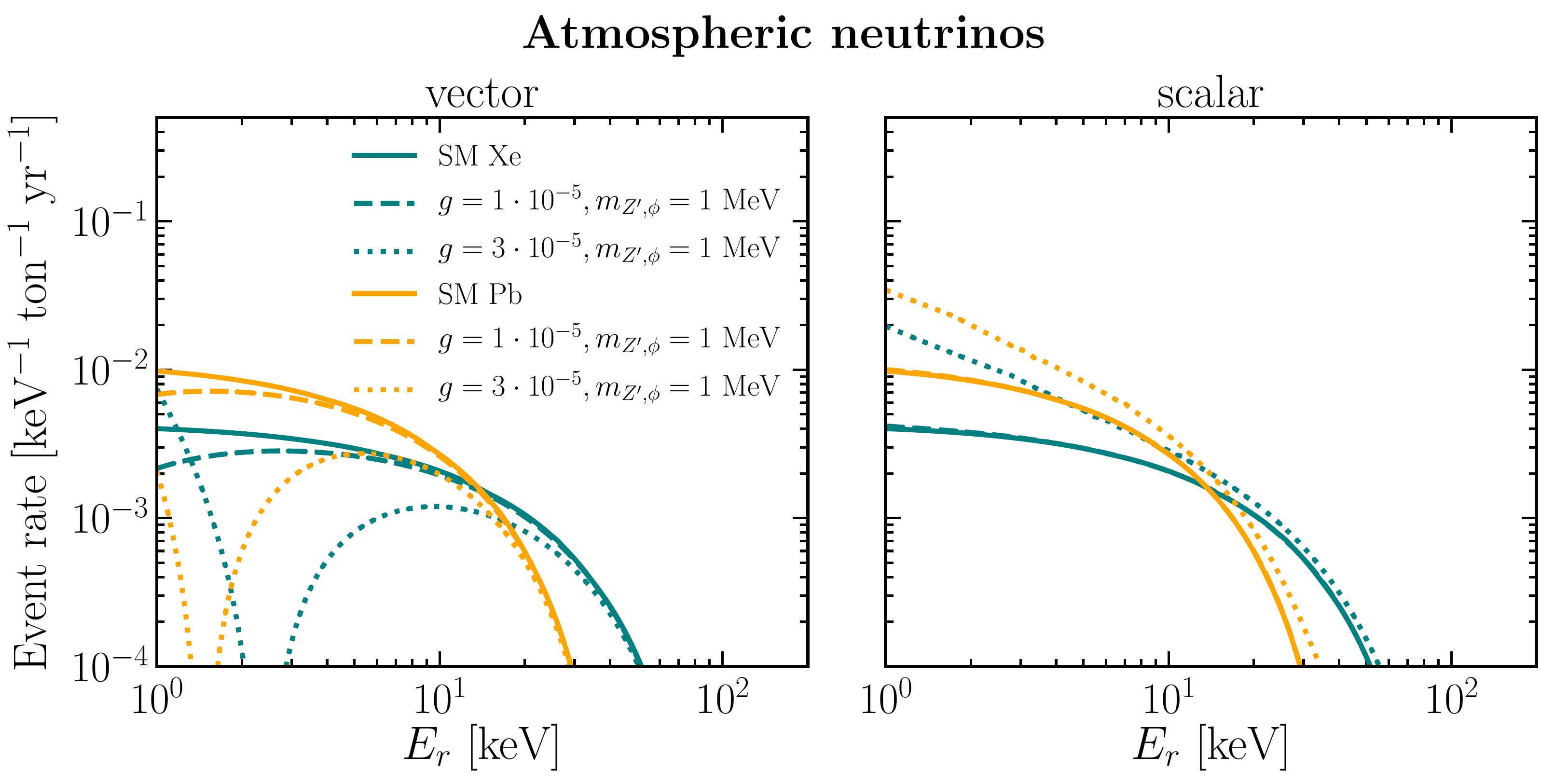}
    \caption{Expected event rate as a function of the nuclear recoil energy induced by atmospheric neutrinos in Pb (orange) and Xe (blue) based detectors. The standard event rates (solid lines) are shown  together with the ones involving non-standard coherent scattering  on the left (right) for vector (scalar) mediators. Given the energy range of atmospheric neutrinos, the standard and non-standard rates are comparable for the vector and scalar scenarios.}
    \label{fig:Atm_BSM}
\end{figure*}

Figure~\ref{fig:Atm_BSM} shows the event rate induced by  atmospheric neutrinos in Xe (green) and Pb based (orange) detectors. The expected number of events is reported for the standard (solid lines) and non-standard case (dashed and dotted lines). Differently from the solar event rate, but similarly to the SN rate, the Pb based detector is characterized by a higher number of events per ton at low recoil energies. Analogously to the SN and solar event rates, the new vector mediator can lead to a suppression or an increase of the event rate; while, only an increase is possible in the scalar case (see Eqs.~\ref{eq:sigma_total_vector} and \ref{eq:sigma_total_scalar}). Furthermore, by comparing the dashed lines with the ones from the solar neutrino event rate (Fig.~\ref{fig:Solar_BSM}), all being calculated for the same mass and mediator coupling, one can see that  the difference between the standard and non-standard event rate is higher for a vector mediator, in the presence of relatively high energy neutrinos in the flux (closer or higher energy than $E_r^\mathrm{max}$ for which the form factor experiences a dip); this is due to the difference in the neutrino energy dependence of the non-standard cross section terms.

\section{Detector constraints on the mass and coupling of the new mediator}
\label{sec:Statistical analysis}
\begin{figure*}[t!]
    \centering
    \includegraphics[scale=0.45]{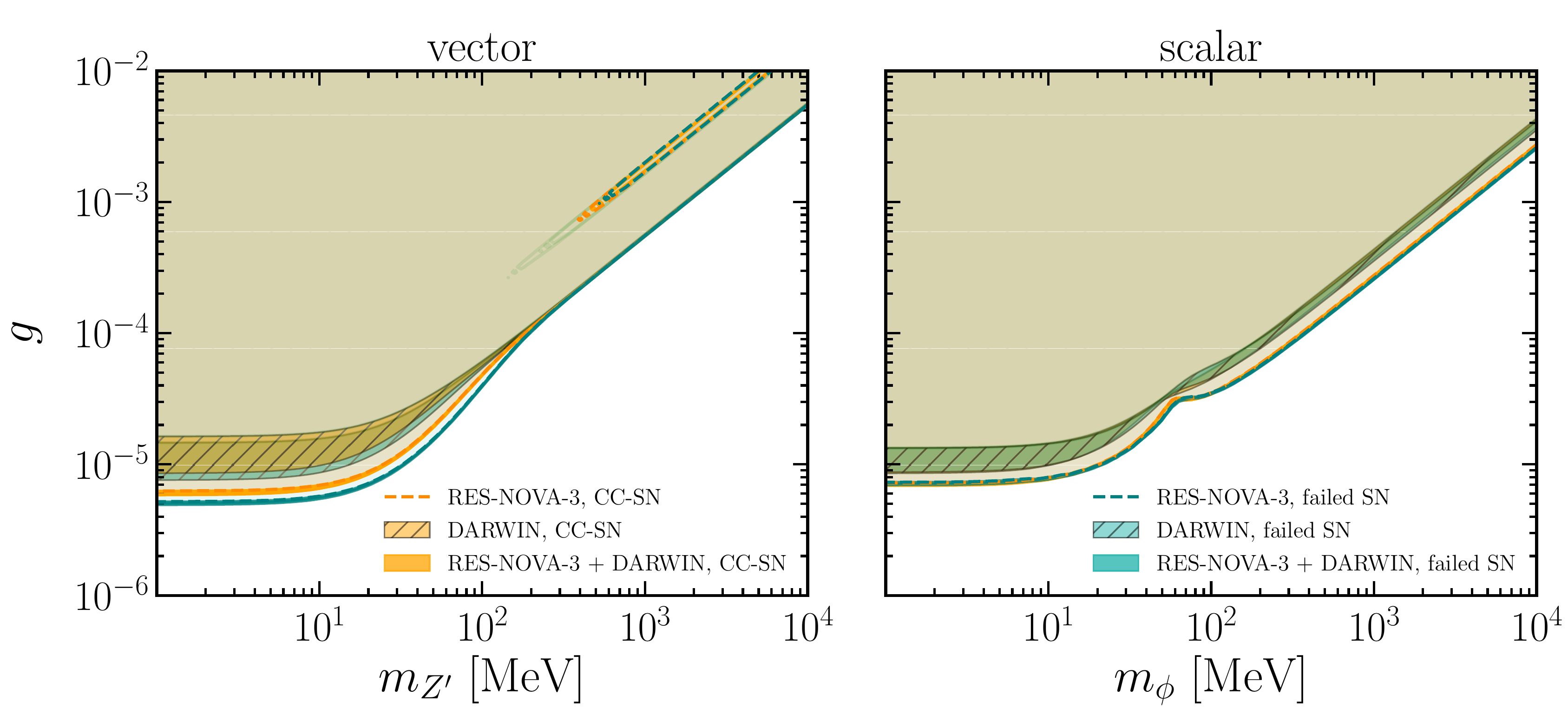}
    \caption{Projected 90$\%$~CL sensitivity bounds on non-standard coherent scattering  in the plane spanned by the mass $m_{Z^\prime,\phi}$ and coupling $g$ of the new mediator. The vector (scalar) case is shown on the left (right) for a SN at $10$~kpc from Earth. The sensitivity bounds for DARWIN (hatched regions), RES-NOVA-3 (dashed lines), and both detectors combined together (solid regions) are reported. The bounds for  core-collapse SNe (CC-SN) are plotted in orange and  the ones for the black hole forming collapses (failed SN) are in blue. The  region of the parameter space excluded in the vector mediator scenario for low and intermediate mediator masses is bigger than the one excluded in the scalar case  because of the more prominent energy dependent features induced by the vector mediator.}
    \label{fig:2D_SN_BMS_10kpc}
\end{figure*}

In this section, we focus on deriving the perspective constraints on the $(g, m_{Z^\prime, \phi})$ parameter space for the new mediator by relying on SN, solar, and atmospheric neutrinos. In order to do that, we implement the $\chi^2$ test using the pull method~\cite{Fogli:2002pt}, with $\chi^2$  defined  as
\begin{equation}
\label{eq:chi_marginalized}
\chi^2 = \underset{x}{\mathrm{min}}{\sum_\mathrm{detector} \left[ \chi^2_{\mathrm{detector}} + \left( \frac{x}{\sigma_{x}} \right)^2 \right]} \ , 
\end{equation}
where $\sigma_x$ is the flux normalization uncertainty.
For each source and for each detector, $\chi^2_\mathrm{detector}$ is 
\begin{equation}
\label{eq:chi_detector}
\chi_\mathrm{detector}^2 = -2 \ln{\frac{L_0}{L_1}} \ .
\end{equation}
The likelihoods are given by
\begin{equation}
\label{eq:L_0}
L_0 = \sum_i P(\lambda = N_{i, {\nu N}}; k = \overline N_{i, \mathrm{SM}}) \ ,
\end{equation}
and 
\begin{equation}
\label{eq:L_1}
L_1 = \sum_i P(\lambda = \overline N_{i, \mathrm{SM}}; k = \overline N_{i, \mathrm{SM}}) \ ,
\end{equation}
where $P$ is the Poisson distribution, $i$ is the bin index and $\overline N_{i,\mathrm{SM}}$ is the number of events observed in the $i$-th bin assuming the standard cross section only, whereas the number of the events when the new mediator is included is  $N_{\nu N} = (1+x) \overline N_{\nu N}$ where $x \; \epsilon \; (-1, \inf)$.

The statistical analysis has been developed by using a $2$~keV binning~\cite{Aalbers:2016jon,Pattavina:2020cqc} of the event rates observed in RES-NOVA-3 and DARWIN. Additionally, the sensitivity bounds rely on a $100 \%$ efficiency for RES-NOVA and DARWIN; for the Xe based detector we also use the energy dependent efficiency of XENON1T~\cite{Aprile:2018dbl,Aprile:2019xxb} for comparison; as such the  excluded regions are defined by a finite band size instead than a curve.

\subsection{Constraints from supernova neutrinos}
\label{sec:Supernova neutrinos results}

By relying on the fact that the new mediator could enhance the coupling of SN neutrinos to matter in the stellar core and, therefore, 
modify the diffusion time of neutrinos, in Fig.~\ref{fig:2D_SN_INSIDE_2} we show the region of the $(g, m)$ parameter space disfavored by this argument.
Additionally, as discussed in Sec.~\ref{sec:SN Effects on the event rates}, SN neutrinos can also undergo non-standard interactions in the detector.

The $\chi^2$ test obtained by using the non-standard recoil rates shown in Fig.~\ref{fig:Rate_SN_BMS}, provides the sensitivity bounds reported in Fig.~\ref{fig:2D_SN_BMS_10kpc} for a SN at $10$~kpc, assuming a normalization uncertainty on the SN signal of $25\%$; this uncertainty takes into account  the early and late time maximal variation in the neutrino signal obtained by comparing different SN hydrodynamical simulations~\cite{OConnor:2018sti}, the SN mass dependence~\cite{Seadrow:2018ftp}, and the uncertainty in  the SN distance determination~\cite{Kachelriess:2004ds,Adams:2013ana} (see Appendix~\ref{Appendix_B} for more details).

The bounds in Fig.~\ref{fig:2D_SN_BMS_10kpc} are obtained for the non-standard vector (on the left) and scalar mediators (on the right) from a galactic SN burst detected in DARWIN (hatched regions), RES-NOVA-3 (dashed lines), and both detectors combined (solid regions). The bounds are shown for the core-collapse (orange) and failed SN cases (blue). 
\begin{figure*}[t!]
    \centering
    \includegraphics[scale=0.45]{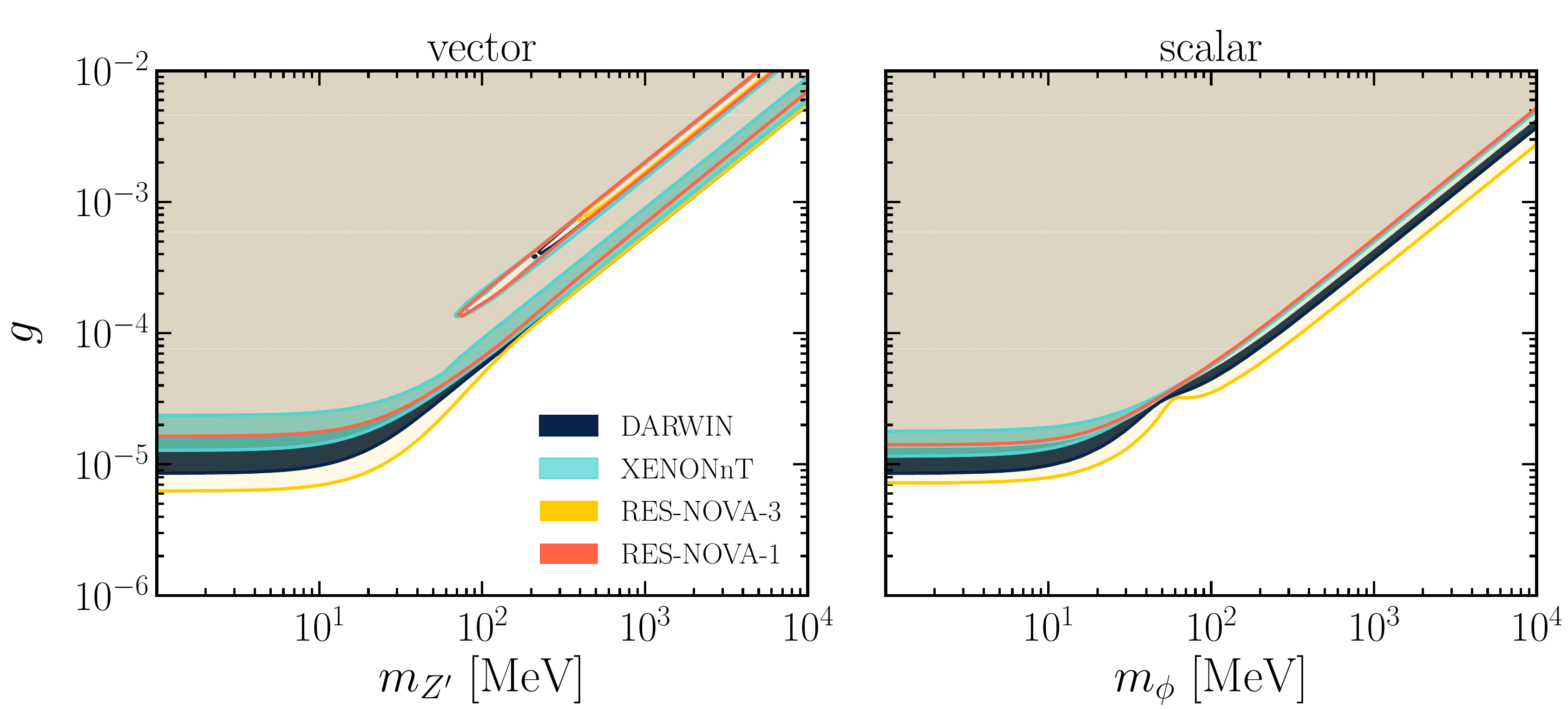}
    \caption{Projected 90$\%$~CL sensitivity bounds on non-standard coherent scattering in the plane spanned by the mass $m_{Z^\prime,\phi}$ and coupling $g$ of the new mediator. The vector (scalar) case is shown on the left (right) panel for a SN at $10$~kpc from Earth. The bounds for DARWIN (navy-blue), XENONnT (cyan) RES-NOVA-3 (yellow) and RES-NOVA-1 (red) are shown, each marginalized over the core-collapse and failed SN models. The bounds are less stringent for smaller detectors. }
    \label{fig:2D_small_big_volume_10kpc}
\end{figure*} 
 
The shape of the excluded region is similar in the vector and scalar mediator scenarios. In the low mass limit, $m_{Z^\prime, \phi} << 2 E_r m_T$, the change in the cross section is caused  by  $g$ only;  this  results in no dependence of the excluded region on the new mediator mass below $\mathcal{O}(10)$~MeV. In the high mass region, $m_{Z^\prime, \phi} << 2 E_r m_T$, the bounds depend  on the effective coupling $g^4 / m_\phi^4$ for the scalar mediator and $g^2/m_{Z^\prime}^2$ ($g^4/m_{Z^\prime}^4$) for the vector mediator when the standard term is comparable to (or much smaller than) the non-standard one in Eq.~\ref{eq:ksi}. In addition, in the high mass limit for the vector mediator case, one can restore the standard cross section when the interference term ($- 3 Q_w {A g^2}/{\sqrt{2}G_F m_{Z^\prime}^2}$) is comparable to the non-standard term in Eq.~\ref{eq:ksi}. This results in a small unconstrained island in the ($g, m_{Z^\prime}$) plane (see Appendix~\ref{Appendix_A} for more details on the event rate dependence on the mediator mass). 
Given that core-collapse SNe and black-hole-forming collapses exhibit a different neutrino signal, we report the exclusion bounds for both  cases in Fig.~\ref{fig:2D_SN_BMS_10kpc}.

 Due to the smaller effective volume of the detector, and smaller event rate per ton (see Fig.~\ref{fig:Rate_SN_BMS}), the DARWIN bounds are less stringent than the ones obtained {for} RES-NOVA-3, independently of the mediator type. The projected limits have a finite size uncertainty band for DARWIN because we use the 100$\%$ detector efficiency and the energy-dependent one. The efficiency band is  wider in the low mass region as the shape of the recoil spectrum can be distinctly changed by the new mediator. The ruled-out region for the vector mediator in the low and intermediate mass region is bigger for the black hole forming collapse  than for the core-collapse SN. This is due to the higher variability of the event rate shape (see Fig.~\ref{fig:Rate_SN_BMS}) in that region than in the case of the large mediator mass, for which only the effective coupling matters; the latter only affects the normalization of the event rate, and the fact that the event rate extends to higher recoil energies due to the hotter spectra of neutrinos from black-hole-forming collapses does not matter anymore. This is not visible for the scalar mediator scenario as the rate can only increase with respect to the standard one.

Given the high event rate foreseen for SN neutrinos, we also explore the sensitivity limits for smaller detector volumes. In particular, we focus on  RES-NOVA-1 ($2.4$~ton)~\cite{Pattavina:2020cqc}, which should be operational within the next $2$ years, and XENONnT ($4$~ton)~\cite{Aprile:2020vtw}---whose tank is currently being filled with liquid Xe. For both detectors, we assume the same thresholds and efficiencies as for the RES-NOVA-3 and DARWIN.
The results are reported in Fig.~\ref{fig:2D_small_big_volume_10kpc};  the sensitivity increases with the effective volume of the detector, as expected. Despite the higher event rate per one ton of Pb (Fig.~\ref{fig:Rate_SN_BMS}), the limits from RES-NOVA-1 are worse than the ones from XENONnT due to the smaller effective  volume of the detector, unless the XENONnT efficiency is worse than 100$\%$. Note that, in this figure, as well as in {Fig.~\ref{fig:BL} and Fig.~\ref{fig:scalar}}, we report the combined bound obtained by marginalizing over the two different SN types (CC-SN and failed SN).


\subsection{Constraints from solar and atmospheric neutrinos}
\label{sec:Solar and atmospheric neutrinos results}
As shown in Secs.~\ref{sec:Solar neutrinos} and \ref{sec:Atmospheric neutrinos}, the recoil energy spectrum  from solar and atmospheric neutrinos has characteristic signatures in the presence of non-standard coherent scattering.  In the following, we assume  that the 1$\sigma$ normalization uncertainty for the atmospheric and solar neutrino fluxes is $\sigma_\mathrm{atm} = 20 \%$~\cite{Honda:2011nf} and $\sigma_\mathrm{sol} = 10 \%$~\cite{Bellini:2008mr,Abe:2016nxk,Anderson:2018ukb}, respectively. {Note that, despite the fact that $\sigma_\mathrm{atm}$ may be larger than $20 \%$ for $E \lesssim 100~\mathrm{MeV}$, due to the increase of the cross section as a function of the energy and the shape of the flux of atmospheric neutrinos, approximately half of the recoil event rate is caused by the atmospheric neutrinos with energies $\gtrsim 100~\mathrm{MeV}$. Hence, in the following, we rely on the most optimistic scenario with $\sigma_\mathrm{atm} = 20\%$.}

\begin{figure*}[t]
    \centering
    \includegraphics[scale=0.45]{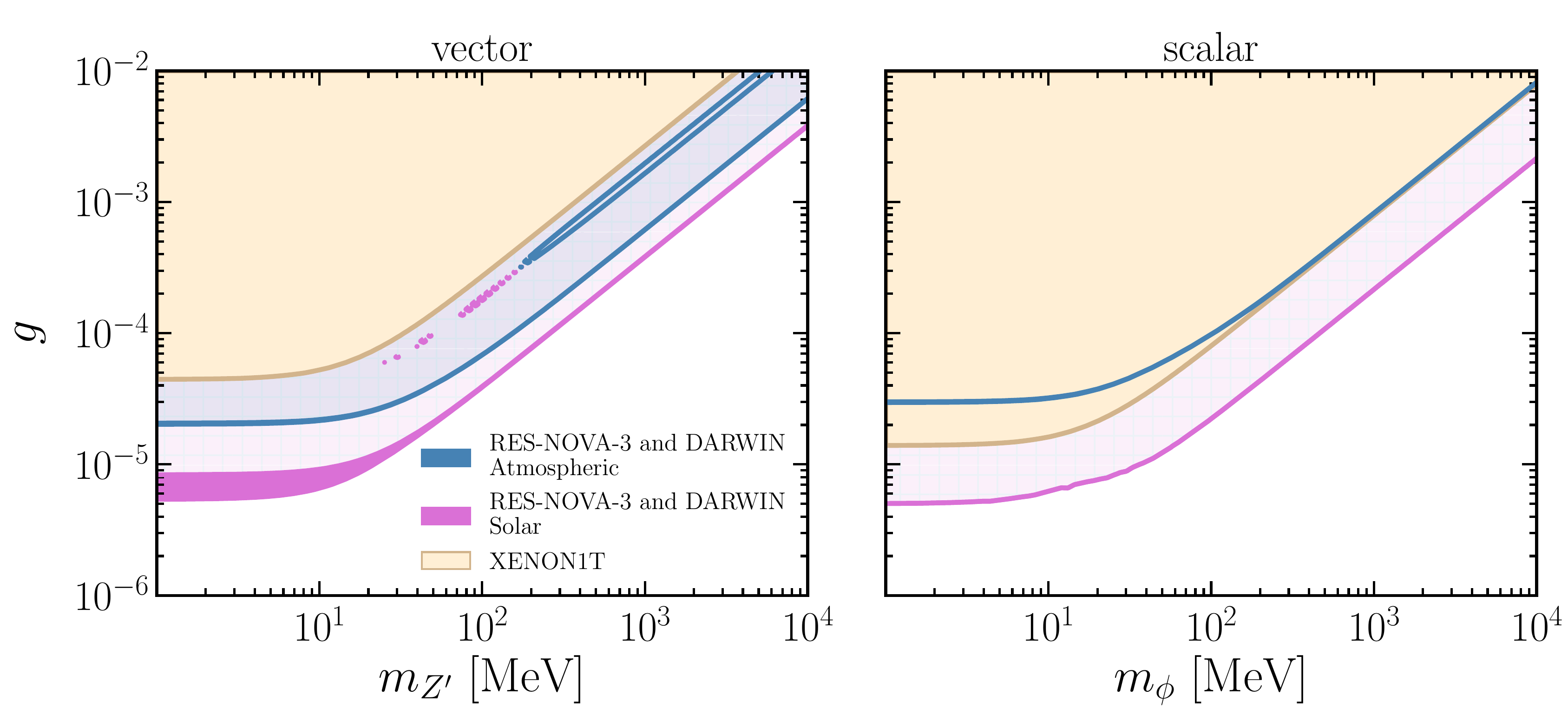}
    \caption{Projected 90$\%$ CL sensitivity bounds on non-standard neutrino-nucleus interactions in the plane spanned by the mass and coupling of the new mediator. The vector (scalar) case is shown on the left (right). The bounds for the atmospheric neutrino signal are plotted in blue and for solar neutrinos in pink. The sensitivity bounds have been derived by relying on the detection of  solar and atmospheric neutrinos in RES-NOVA-3 and DARWIN for $1$~yr exposure. In addition, the bounds obtained using the XENON1T limits on WIMPs~\cite{Aprile:2019xxb} are also shown in beige. Due to the much higher even rate of solar neutrinos, the solar bounds are more stringent than the ones coming from atmospheric neutrinos.}
    \label{fig:2D_SUN_ATM}
\end{figure*}
Figure~\ref{fig:2D_SUN_ATM} shows the projected sensitivity bounds  on the mass and  coupling of the new mediator coming from the detection of solar and atmospheric neutrinos in DARWIN and RES-NOVA-3 for $1$~yr exposure.
Figure~\ref{fig:2D_SUN_ATM} shows that the bounds from atmospheric neutrinos are less stringent than the ones from solar neutrinos. In fact, by comparing Figs.~\ref{fig:Solar_BSM} and \ref{fig:Atm_BSM}, one can see that the number of events induced by solar neutrinos is much larger than  in the case of  atmospheric neutrinos, despite the recoil spectra extending to higher energies for the latter.
However, for the vector mediator (left panel), as the mediator mass increases, the bounds from solar neutrinos become slightly less competitive compared to the ones from atmospheric neutrinos. This is due to the possibility of experiencing energy dependent features over a larger recoil range than in the case of solar neutrinos. Nevertheless, due to the much smaller total number of events for atmospheric neutrinos, the solar bounds are always better.  
The  solar neutrino bounds on the scalar mediator are better than the ones on  the vector mediator, and the opposite effect is true for atmospheric neutrinos. This is due to the fact that atmospheric neutrinos have much higher energies than solar neutrinos, and  the non-standard cross section for the scalar mediator is smaller than the one for the vector mediator case in the high-energy neutrino limit (see Sec.~\ref{sec:Atmospheric neutrinos}).
Since the volume of RES-NOVA-3 is expect to be $10$ times larger  than the one of DARWIN, there is no uncertainty band connected to the  different efficiencies used to calculate the bounds except for the solar neutrino vector case, where the Xe based detector probes  twice the energy range than  the Pb one. This is not evident for the scalar case due to the insensitivity of the low mediator mass cross section to the target material (see Sec.~\ref{sec:Solar neutrinos}).

Our projected 90$\%$CL sensitivity for solar neutrinos in RES-NOVA-3 and DARWIN in the vector mediator case is in good agreement with the one in Ref.~\cite{Cerdeno:2016sfi}, where the authors calculated the bounds for the Xe based detector using $200$~ton-yrs exposure, 100$\%$ detector efficiency, $1$~keV energy threshold, and zero background. However, the $200$~ton~yrs exposure in~\cite{Cerdeno:2016sfi} is obtained by using the DARWIN volume integrated over $5$~yrs, while we consider $1$~yr only. The reduced exposure time makes our results a bit less competitive than the ones reported in Ref.~\cite{Cerdeno:2016sfi}, despite the fact that we combine RES-NOVA-3 and DARWIN. Our bounds are in fact completely driven by the Xe based detector that performs better in the energy range of solar neutrinos and are not improved much by the addition of RES-NOVA-3.

{
For completeness, despite the fact that atmospheric neutrinos  are not competitive with solar or SN neutrinos in constraining non-standard  mediators, we have investigated how the atmospheric neutrino bounds  change if  $\sigma_\mathrm{atm}=50\%$. In this case, the limits for the vector boson, obtained for RES-NOVA-3 and DARWIN, relax by $\sim20\%$ while the ones for the scalar mediator shift marginally (results not shown here). We identify a  trend similar to the one found for the SN neutrino flux normalization uncertainty (see Appendix~\ref{Appendix_A}). 
}

Additionally, we estimate the current $90\%$~CL upper constraints on the mass and coupling of the new mediator of non-standard neutrino-nucleus interactions from the XENON1T data~\cite{Aprile:2019xxb}, assuming the current detector threshold ($0.7$~keV) and efficiency for the S2 channel. In order to do that, we estimate the maximal event rate per $1$ year of XENON1T exposure by taking the upper bound coming from calculating the event rates for the correspondent bounds on the WIMP masses and spin independent cross sections through the  WIMPrates package~\cite{jelle_aalbers_2019_3551727}, under the assumption of a standard isothermal dark matter halo \cite{Aprile:2019xxb,Aprile:2018dbl}. The resultant bounds for vector and scalar mediators are presented in Fig.~\ref{fig:2D_SUN_ATM}. Similarly to the bounds obtained by looking at the diffusion time for SN neutrinos, the bounds for the scalar mediator are more stringent.

{
\section{Discussion}
\label{sec:Discussion}

In this paper, we focus on the generic scenario involving the coupling of the new mediator to neutrinos and quarks with $g = \sqrt{g_\nu g_q}$. In this Section, first we briefly summarize our findings and then discuss how the bounds obtained in this work map onto two non-standard mediator models widely adopted in the literature. We focus on the gauge $U(1)_\mathrm{B-L}$ model proposed in Ref.~\cite{Harnik:2012ni} and on a more generic case of a scalar mediator coupling to neutrinos and quarks.

\subsection{Summary of our findings}
\label{sec:summary}

\begin{figure}[t!]
 \centering
 \includegraphics[width=\columnwidth]{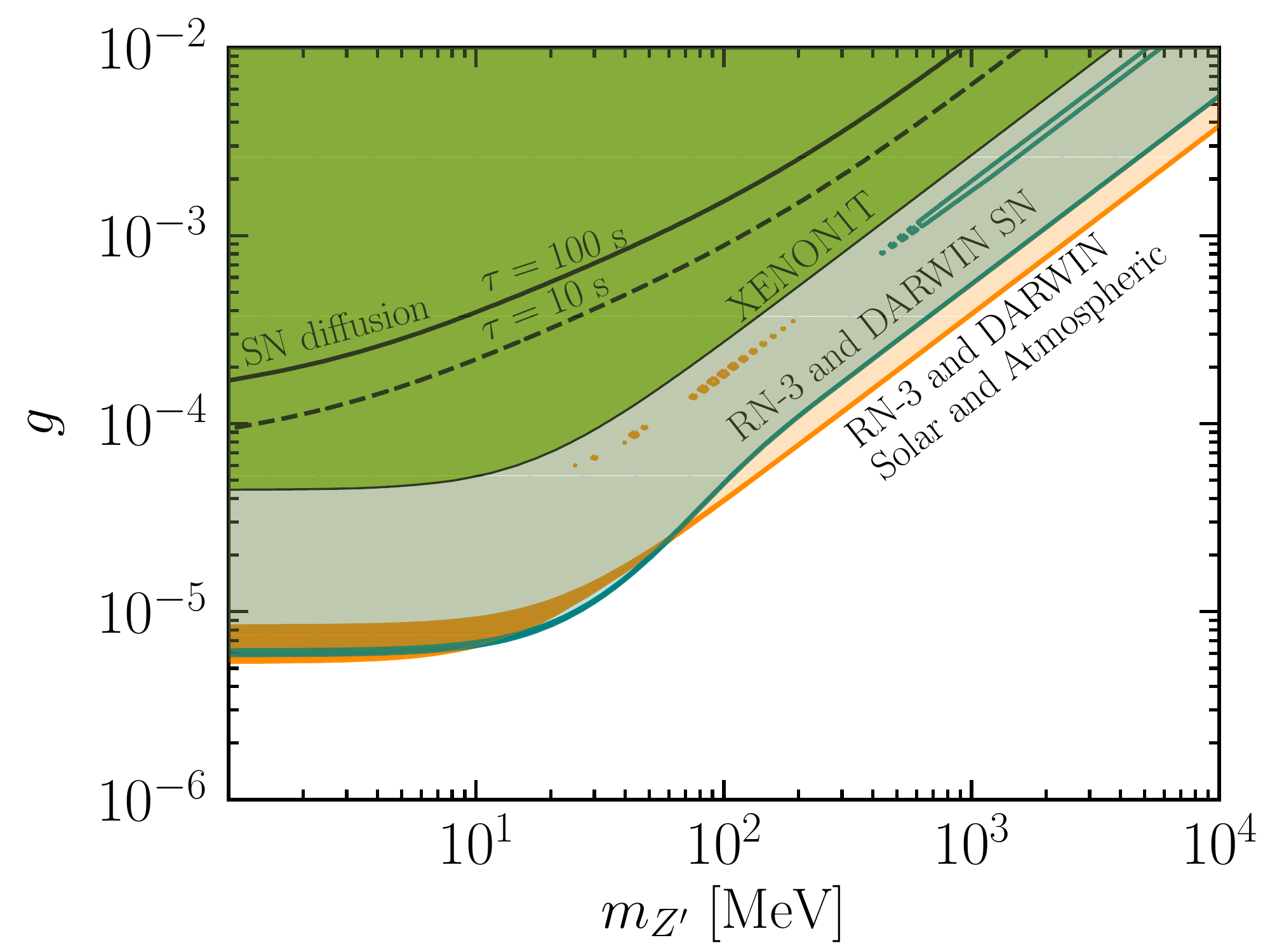}
 \caption{Summary of the bounds derived in this work on the new vector mediator coupling in the plane spanned by the vector mass $m_{Z^\prime}$ and coupling $g$. Our new sensitivity bounds come from considering non-standard neutrino-nucleus (nucleon) interactions in the SN core (marked by solid and dashed black lines), by detecting a neutrino burst from a galactic SN (green line and hatched region), as well as $1$~yr exposure to solar and atmospheric neutrinos (orange line and hatched region) in DARWIN and RES-NOVA-3 (RN-3). The sensitivity of XENON1T has been calculated by relying on the limits provided in Ref.~\cite{Aprile:2019xxb} (light green). DARWIN and RES-NOVA-3 have the potential to exclude the largest region of the parameter space. The bounds plotted here are for a vector mediator; similar ones have been derived for a scalar mediator.
}
\label{fig:Fig0}
\end{figure}

Figure~\ref{fig:Fig0} displays the parameter space spanned by the new  vector mediator coupling $g$ and mass $m_{Z^\prime}$ excluded in this work by relying on the neutrino burst expected from a galactic SN, as well as solar and atmospheric neutrinos, together with the exclusion bounds for XENON1T~\cite{Aprile:2019xxb} calculated in this work. We refrain from showing the analogous plot for the  scalar mediator, since the bounds are very similar to the vector mediator scenario.

Our work shows that the detection of SN neutrinos or $1$~yr exposure to solar and atmospheric neutrinos would allow to probe a large region of the parameter space spanned by the mass and coupling of the new mediator. Under the assumption of optimal background tagging, the observation of solar, atmospheric or SN neutrinos in RES-NOVA-3 and DARWIN gives similar sensitivity to New Physics.

\subsection{Comparison with existing bounds: vector mediator for the $U(1)_\mathrm{B-L}$ model}
\label{sec:BL}

Here we discuss the constraints that apply to the mass and the coupling of the new vector mediator for the $U(1)_\mathrm{B-L}$ gauge boson proposed in Ref.~\cite{Harnik:2012ni}, with coupling to quarks $g_q = 1/3 g_\mathrm{B-L}$ and leptons ($l$) $g_l = g_\nu = -g_\mathrm{B-L}$. A summary of the constrained region of the parameter space is reported in Fig.~\ref{fig:BL}.

\begin{itemize}
\item \textbf{Non-standard coupling to quarks only.}
Constraints on the non-standard coupling to nucleons or quarks (beige, right-slash hatched regions in Fig.~\ref{fig:BL}) to the new mediator can be split into two categories: terrestrial experiments and astrophysical limits. Examples of the former come from the pion decay experiments 
($\pi$ decay)~\cite{Dobroliubov:1987cb,Dobroliubov:1990ye,Gninenko:1998pm} and neutron scattering on the $^{208}$Pb target (n-Pb)~\cite{Barbieri:1975xy,Leeb:1992qf,Schmiedmayer:1988bm}.
As for astrophysical constraints, one can consider the nucleon-nucleon Bremsstrahlung as an additional source of SN cooling~\cite{Grifols:1988fv,Rrapaj:2015wgs} (SN 1987A~$q$), and the impact of non-standard interactions between protons on the Coulomb barrier penetration in the sun~\cite{Suliga:2020lir} (Sun pp).

\item\textbf{Non-standard coupling to neutrinos only.}
Constraints on non-standard mediators coupling to neutrinos are plotted in beige as left-slash hatched regions in Fig.~\ref{fig:BL}. 
These bounds have been derived by  looking at the possible effects of the non-standard mediator on the decay of $W$ and $K$~\cite{Laha:2013xua} ($W$ decay, $K$ decay).
Non-standard interactions  could also cause visible effects on high-energy  neutrinos of astrophysical origin. If the high-energy neutrinos interact with the relic neutrinos (cosmic neutrino background, see, e.g., Ref.~\cite{Vitagliano:2019yzm}) via the exchange of a non-standard mediator, spectral distortions or delays should be expected in the signal observable at Earth~\cite{Blum:2014ewa,Ioka:2014kca,Ibe:2014pja,DiFranzo:2015qea,Murase:2019xqi,Bustamante:2020mep}. In Ref.~\cite{Bustamante:2020mep}, a  statistical analysis has been performed to search for signs of non-standard interactions among neutrinos in the diffuse flux of high-energy  neutrinos detected by the IceCube Neutrino Observatory by relying on the  High Energy Starting Events~(IC HESE). 
By using similar arguments, another independent constraint has been reported in Ref.~\cite{Kelly:2018tyg} by exploiting the possible detection of high-energy neutrinos from the blazar TXS 0506+056 (IC TXS).

The region of the parameter space disfavored by non-standard interactions between neutrinos coming from the SN 1987A was studied in Refs.~\cite{Kolb:1987qy,Shalgar:2019rqe}. In this case, limits were placed by considering non-standard interactions of SN neutrinos with relic neutrinos for mediator masses $m_\mathrm{B-L} \lesssim 0.1~\mathrm{MeV}$. 
Additionally, Ref.~\cite{Shalgar:2019rqe} examined the consequences of non-standard neutrino-neutrino interactions occurring in the SN core on the delayed neutrino heating mechanism~\cite{Bethe:1984ux}. The region of the parameter space disfavored by this argument is shown in Fig.~\ref{fig:BL} (SN 1987A $\nu$).

The impact of the non-standard vector mediator on the Big Bang Nucleosynthesis (BBN) has been discussed in Refs.~\cite{Ahlgren:2013wba,Huang:2017egl}.
The Boltzmann equations in the isotropic and homogeneous Universe have been solved in the presence of non-standard interactions. The parameter space for which the change in the number of the effective relativistic degrees of freedom  is such that $\Delta N_\mathrm{eff} > 1$ in the non-standard scenario has been excluded; this was done by translating the obtained $\Delta N_\mathrm{eff}$ in the change in the primordial abundance of deuterium, and mass fraction of helium. The limit (BBN) in Fig.~\ref{fig:BL} comes from Ref.~\cite{Huang:2017egl}.

The existence of a new vector mediator might also  impact the cosmic microwave background~(CMB) through the effect of increasing the power on small scales in the Planck data~\cite{Archidiacono:2013dua,Cyr-Racine:2013jua}. This limit is indicated as CMB $\nu$ in Fig.~\ref{fig:BL}.

\end{itemize}
\begin{figure}[t!]
 \centering
 \includegraphics[width=\columnwidth]{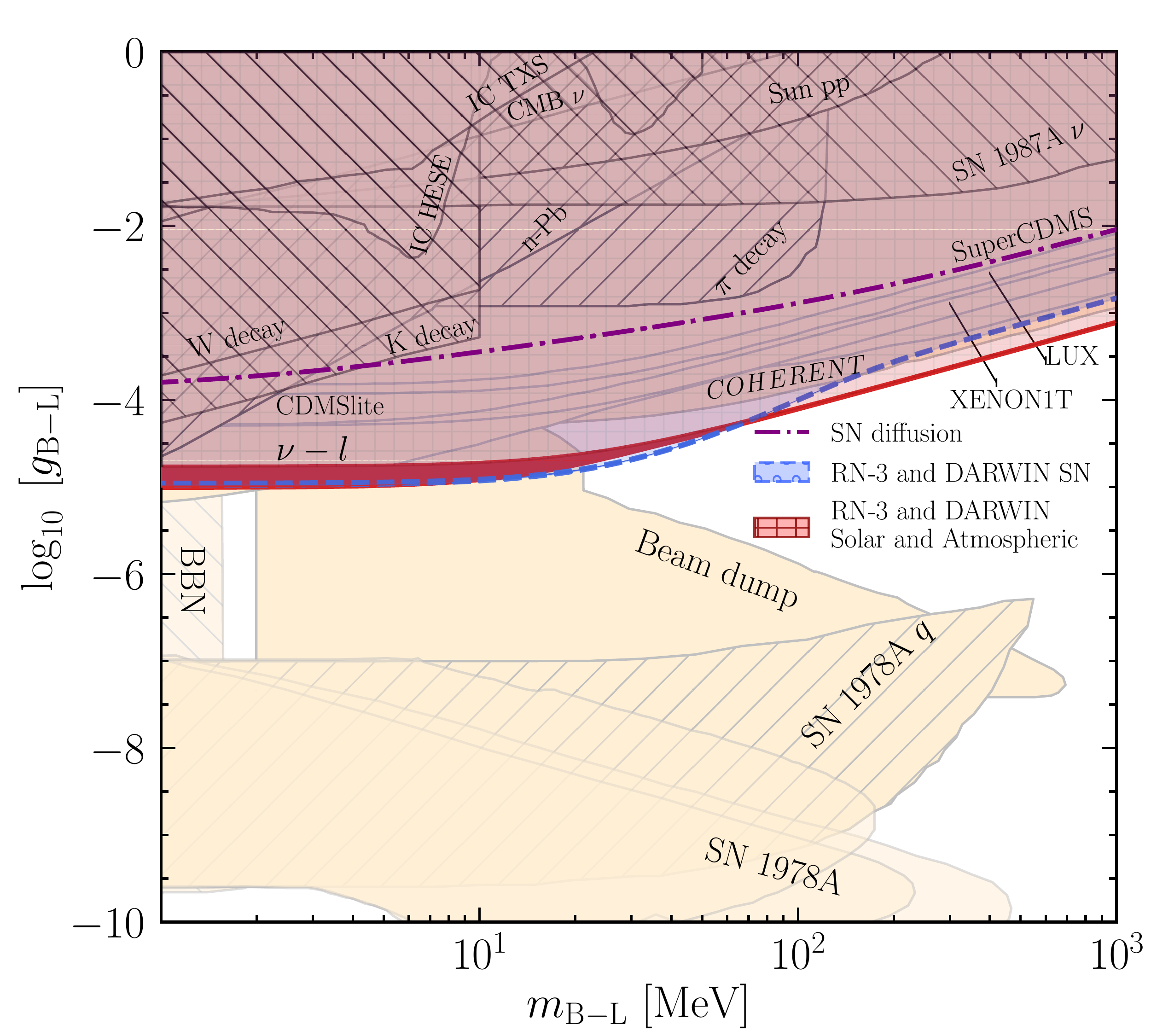}
 \caption{
 Bounds on the $U(1)_\mathrm{B-L}$ model in the plane spanned by the vector mediator mass $m_\mathrm{B-L}$ and coupling $g_\mathrm{B-L}$. Our new sensitivity bounds come from considering non-standard neutrino-nucleus (nucleon) interactions in the SN core (marked by purple dash-dotted line), by detecting a neutrino burst from a galactic SN (blue dashed line and hatched region), as well as $1$~yr exposure to solar and atmospheric neutrinos (red solid line and hatched region) in DARWIN and RES-NOVA-3 (RN-3). Our prospective limits are contrasted with various constraints.
 The limits which apply  to quarks only (shaded, right slash hatch): SN 1987A~\cite{Rrapaj:2015wgs} (SN 1987A $q$), neutron-Pb scattering~\cite{Farzan:2016wym} (n-Pb), $\pi$ decays experiments ($\pi$ decay), and Coulomb barrier penetration inside the Sun (Sun pp)~\cite{Suliga:2020lir}.
 We also show the bounds that require coupling of the non-standard mediator to neutrinos only (beige, right-slashed hatched regions): Big Bang Nucleosynthesis~\cite{Huang:2017egl} (BBN), SN 1987A~\cite{Shalgar:2019rqe}(SN 1987A $\nu$), K and W decays~\cite{Laha:2013xua}, high-energy astrophysical neutrinos (IC HESE)~\cite{Bustamante:2020mep}, neutrinos from the TXS 0506+056 source~\cite{Kelly:2018tyg} (IC TXS), and Cosmic Neutrino Background (CMB)~\cite{Archidiacono:2013dua}.
 The bounds which apply to leptons and quarks (beige, solid regions): COHERENT~\cite{Cadeddu:2020nbr}, SuperCDMS, CDMSlite, and LUX data~\cite{Cerdeno:2016sfi}, beam dump experiments~\cite{Bauer:2018onh}, neutrino-lepton scattering~\cite{Harnik:2012ni,Lindner:2018kjo} ($\nu-l$), and SN 1987A~\cite{Chang:2016ntp,Croon:2020lrf}. 
 The sensitivity of XENON1T has been calculated by relying on the limits provided in Ref.~\cite{Aprile:2019xxb} (beige, solid region).
 DARWIN and RES-NOVA-3 have the potential to exclude a not yet probed region of the parameter space for high $m_\mathrm{B-L}$.
}
\label{fig:BL}
\end{figure}
\begin{itemize}
\item\textbf{Non-standard coupling to charged leptons and any coupling.}
The coupling of the new mediator to  active neutrinos and charged leptons allows to use the data from neutrino experiments able of observing $\nu + l \rightarrow \nu +l$, such as 
Borexino~\cite{Bellini:2011rx}, Texono~\cite{Deniz:2009mu}, GEMMA~\cite{Beda:2009kx}, and CHARM-II~\cite{Vilain:1993kd,Vilain:1994qy} to put bounds on the mass and coupling of the mediator. In Fig.~\ref{fig:BL}, bounds coming from these experiments are marked with $\nu-l$~\cite{Harnik:2012ni,Lindner:2018kjo,Bauer:2018onh,Cadeddu:2020nbr}.

We also show the bound coming from the beam dump experiments (Beam dump). One can distinguish the electron case, where only the coupling to electrons is considered. Examples of these experiments are Fermilab~E774~\cite{Bross:1989mp}, SLAC~E137, SLAC~E141~\cite{Bjorken:1988as,Bjorken:2009mm}, and 
Orsay~\cite{DAVIER1989150}. For the proton beam dumps--LSND~\cite{Athanassopoulos:1997er} and U70/NuCAL~\cite{Blumlein:2011mv,Blumlein:2013cua}--both quarks and leptons need to couple to the non-standard mediator. The bounds in  Fig.~\ref{fig:BL} include all the mentioned experiments and have been taken from Ref.~\cite{Bauer:2018onh}. We also note that one can constrain the  $U(1)_\mathrm{B-L}$ model from fixed target experiments, such as NA48/2~\cite{Batley:2015lha}, APEX~\cite{Abrahamyan:2011gv}, A1/MAMI~\cite{Merkel:2011ze,Merkel:2014avp}, and NA64~\cite{Banerjee:2016tad} (see, e.g., Refs.~\cite{Bauer:2018onh,Cadeddu:2020nbr}). However, these experiments do not exclude any region of the parameter space already constrained by other arguments reported here, hence we refrain from showing them explicitly.

The bounds on the $U(1)_\mathrm{B-L}$ model from the SN 1978A where the mediator couples to neutrinos, charged leptons, and quarks have been studied in Ref.~\cite{Chang:2016ntp,Croon:2020lrf} (SN 1987A). There the bounds have been derived by relying on a static hydrodynamical background at $1$~s post bounce from a one-dimensional SN simulation  and by applying the cooling criterion~\cite{Raffelt:1996wa}. Various processes were included, however, it has been found that the inverse decay of $\nu + \bar\nu \rightarrow Z^\prime$ and the semi-Compton scattering contributed the most to the derived limits. Interestingly, a SN simulation taking into account muon production has been considered in Ref.~\cite{Croon:2020lrf}; this led to a shift of the earlier bounds~\cite{Chang:2016ntp}.

The current limits from COHERENT~\cite{AristizabalSierra:2019ykk,Cadeddu:2020nbr}, Super-CDMS, CDMSlite and LUX data~\cite{Cerdeno:2016sfi,Agnese:2015nto,Agnese:2014aze,Akerib:2015rjg} are also shown in Fig.~\ref{fig:BL}. In addition, we also show the exclusion bounds for XENON1T~\cite{Aprile:2019xxb} calculated in Sec.~\ref{sec:Statistical analysis}.

\end{itemize}
Our work shows that the detection of SN neutrinos, or $1$~yr exposure to solar and atmospheric neutrinos, would allow to probe the largest region of the parameter space spanned by the mass and coupling of the new mediator among the experiments capable of observing C$\nu$NS. Under the assumption of optimal background tagging and in the limit of low mediator mass, the potential improvement with respect to the limits from COHERENT~\cite{AristizabalSierra:2019ykk,Cadeddu:2020nbr}, which are currently the most stringent ones, is up to 50$\%$.
In addition, our bounds on the $U(1)_\mathrm{B-L}$ model promise to place most sensitive bounds on a fraction of the parameter space (20~MeV$ \lesssim m_\mathrm{B-L} \lesssim 1~\mathrm{GeV}$) yet unconstrained by other, listed in this section limits.

\subsection{Comparison with existing bounds: scalar mediator}
\label{sec:Scalar_model}

We will now discuss bounds that apply for the scalar coupling to neutrinos and quarks (Fig.~\ref{fig:scalar}). We assume that $g_\nu = g_q = g_\phi$. For examples of specific models we refer the reader to, e.g., Refs.~\cite{Farzan:2018gtr,Berryman:2018ogk}.  

\begin{figure}[t!]
 \centering
 \includegraphics[width=\columnwidth]{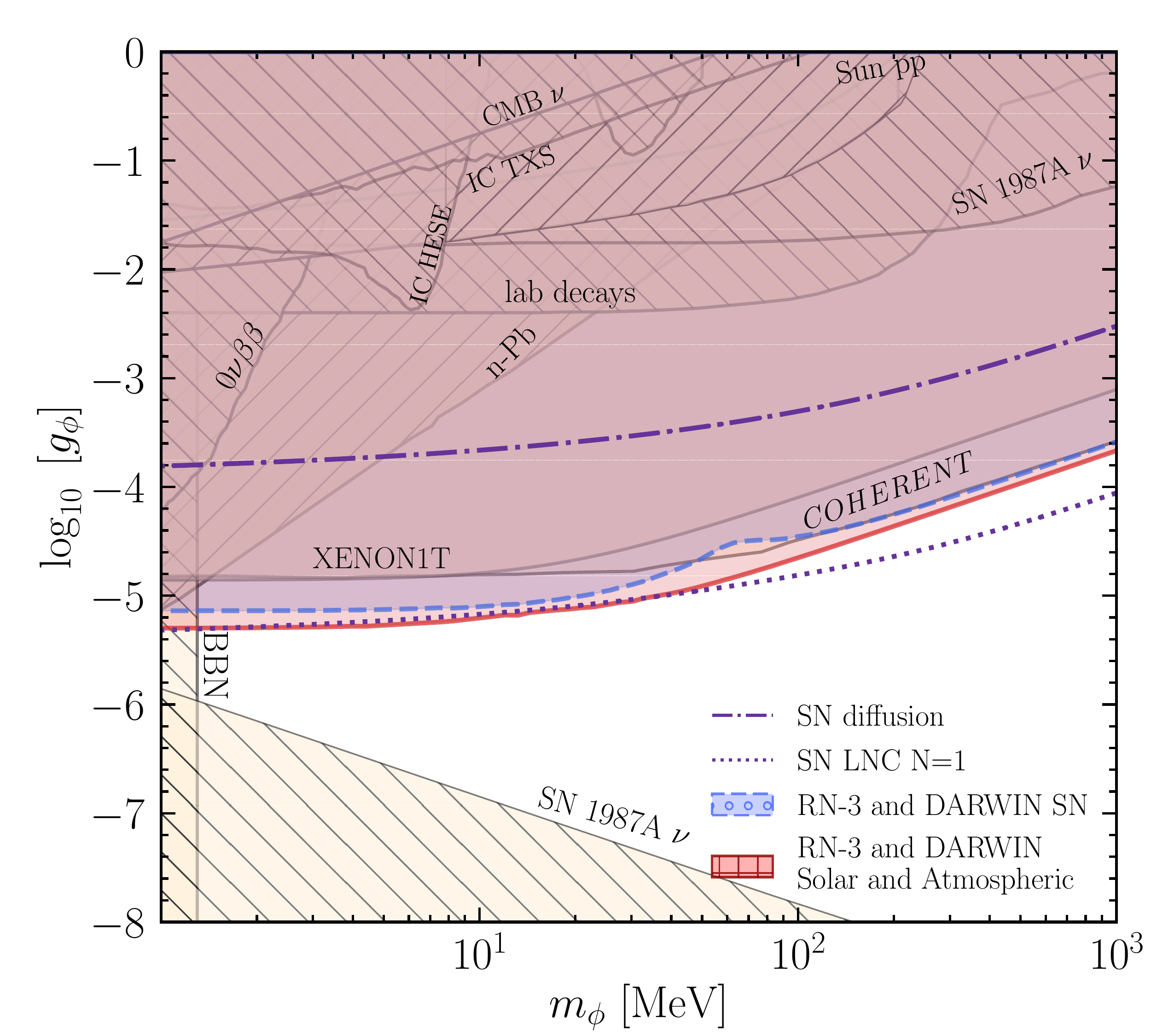}
 \caption{
 Bounds on the scalar coupling to neutrinos and quarks in the plane spanned by the mediator mass $m_\phi$ and coupling $g_\phi$. Our new sensitivity bounds come from considering non-standard neutrino-nucleus (nucleon) interactions in the SN core (marked by purple dash-dotted and dotted lines), by detecting a neutrino burst from a galactic SN (blue dashed line and hatched region), as well as $1$~yr exposure to solar and atmospheric neutrinos (red solid line and hatched region) in DARWIN and RES-NOVA-3 (RN-3). Our prospective limits are contrasted with other constraints.
 The limits which apply only to quarks (shaded, right slash hatch): neutron-Pb scattering~\cite{Farzan:2016wym} (n-Pb), and Coulomb barrier penetration in the sun (Sun pp)~\cite{Suliga:2020lir}.
 We also show the bounds  only requiring coupling of the non-standard mediator to neutrinos (beige, right-slashed hatched regions): Big Bang Nucleosynthesis~\cite{Blinov:2019gcj} (BBN), SN 1987A~\cite{Shalgar:2019rqe}(SN 1987A $\nu$, top region) and ~\cite{Heurtier:2016otg}(SN 1987A $\nu$, bottom region), high-energy astrophysical neutrinos (IC HESE)~\cite{Bustamante:2020mep}, neutrinos from the TXS 0506+056 source~\cite{Kelly:2018tyg} (IC TXS), and Cosmic Neutrino Background (CMB)~\cite{Cyr-Racine:2013jua}.
 The bounds which apply to neutrinos and quarks (beige, solid regions): COHERENT~\cite{AristizabalSierra:2019ykk}.
 The sensitivity of XENON1T has been calculated by relying on the limits provided in Ref.~\cite{Aprile:2019xxb} (beige, solid region).
 Likewise for the $U(1)_\mathrm{B-L}$ model, DARWIN and RES-NOVA-3 have the potential to exclude the not yet probed region of the parameter space for high $m_\phi$.
}
\label{fig:scalar}
\end{figure}

\begin{itemize}
\item{\textbf{Non-standard coupling to quarks.}
The bounds that apply to the scalar mediator coupling to quarks based on the same arguments as for the vector mediator discussed in Sec.~\ref{sec:BL} are: neutrons scattering on the $^{208}$Pb target (n-Pb)~\cite{Barbieri:1975xy,Leeb:1992qf,Schmiedmayer:1988bm}, nucleon-nucleon Bremsstrahlung in stars~\cite{Grifols:1988fv} (for $m_\phi<$ 2~keV, therefore not shown on Fig.~\ref{fig:scalar}), and changes of the Coulomb barrier penetration in the sun~\cite{Suliga:2020lir} (Sun pp).
}

\item{\textbf{Non-standard coupling to neutrinos.}
Laboratory experiments focused on the precision measurements of $\tau$, mesons, and $Z$ decay widths~\cite{Bilenky:1999dn,Lessa:2007up,Pasquini:2015fjv,Berryman:2018ogk,Krnjaic:2019rsv,Brdar:2020nbj} can set limits on the mass and the coupling of the scalar mediator coupling to neutrinos (lab decays). 
Double-beta decay experiments (see, e.g., recent results from GERDA \cite{Agostini:2015nwa}) can also place constraints on the mass and coupling of the new LNV scalar mediator. For scalar mediator masses below the $Q$-value of certain nuclei that could undergo double beta decay, new double-beta decay channels might open up~\cite{Rodejohann:2011mu,Gando:2012pj,Blum:2018ljv,Berryman:2018ogk,Deppisch:2020sqh} ($0\nu\beta\beta$).  

Moreover, similar to the vector mediator case, bounds on the mass and coupling of the non-standard mediator from the CMB and BBN \cite{Cyr-Racine:2013jua,Escudero:2019gvw,Huang:2017egl,Oldengott:2017fhy,Barenboim:2019tux,DiValentino:2017oaw,Blinov:2019gcj,Grohs:2020xxd,Huang:2021dba} (BBN), and high-energy neutrinos seen by IceCube~\cite{Kelly:2018tyg,Ng:2014pca,Bustamante:2020mep} (IC HESE, IC TXS) apply as well.
The SN 1987A constraints on the scalar mediator  follow the same argument as for the vector case~\cite{Kolb:1987qy,Heurtier:2016otg,Shalgar:2019rqe} (SN 1987A $\nu$). However, due to the helicity suppression, these limits should be interpreted as the ones on a effective coupling which already incorporates the suppression factor.
}

\item{
\textbf{Non-standard coupling to neutrinos and quarks.}
Likewise in the vector mediator case, one can obtain bounds on the scalar mediator coupling to neutrinos and quarks from COHERENT~\cite{AristizabalSierra:2019ykk}, and the direct detection dark matter experiments, such as Super-CDMS, CDMSlite and LUX~\cite{Agnese:2015nto,Agnese:2014aze,Akerib:2015rjg}. We refrain from showing the latter ones, as alike the case of a vector mediator, the limits are expected to be less stringent than the ones from COHERENT and XENON1T.}
\end{itemize}

Analogously as in the vector mediator model case, our limits have the potential to rule out a yet unconstrained region of the parameter space in the plane spanned by the mass and the coupling of the new scalar mediator. Excitingly, our limits from XENON1T for $m_\phi \lesssim 10~\mathrm{MeV}$ seem to be comparable with the ones from COHERENT~\cite{AristizabalSierra:2017joc}. Intriguingly, the $1$~yr exposure to solar and atmospheric neutrinos or the observation of a galactic SN burst in RES-NOVA and DARWIN would improve the limits by up to $\sim65\%$ compared to COHERENT~\cite{AristizabalSierra:2019ykk}.
We note that the possibility of non-negligible kinetic mixing between the non-standard mediators and their SM counterparts may affect the reported constraints, see e.g., Refs.~\cite{Davoudiasl:2012ag,Abdullah:2018ykz,Croon:2020lrf}.

}

\section{Conclusions}
\label{sec:Conclusions}

We employ astrophysical neutrinos to constrain non-standard coherent neutrino-nucleus scattering for vector and scalar mediators, as summarized in {Figs.~\ref{fig:Fig0}--\ref{fig:scalar}}. In particular, we rely on the impact that the non-standard coherent neutrino-nucleus scattering  would have on the physics of core-collapse supernovae, as well as the effect of non-standard reactions occurring between supernova, solar and atmospheric neutrinos and the nuclei in RES-NOVA, DARWIN, XENONnT, and XENON1T. 

If non-standard coherent neutrino-nucleus scattering should occur in the supernova core, neutrinos would need more time to diffuse out, with possible implications on the supernova explosion mechanism.
In addition, for the lepton number conserving interactions mediated by a scalar particle, we have also calculated the limits on the mass and coupling of the mediator using the cooling criterion~\cite{Raffelt:1996wa}.

In order to derive bounds on non-standard coherent neutrino-nucleus scattering in upcoming CE$\nu$NS detectors, we focus on the fact that the nuclear recoil spectrum changes its shape and normalization in the presence of a new vector or scalar mediator. We have shown that the observation of the neutrino signal from a galactic supernova in DARWIN and RES-NOVA can potentially place the most stringent constraints. The improvement on current bounds from COHERENT~\cite{AristizabalSierra:2019ykk} is up to 50$\%$ in the limit of low mediator masses. Recently Ref.~\cite{Lenardo:2019fcn} highlighted the possibility of reducing the energy threshold in  dual phase xenon detectors to $0.3$~keV. In this case, our bounds would improve significantly (up to $\sim 50\%$).

The sensitivity projections for solar and atmospheric neutrinos show that solar neutrinos will be driving the bounds on the non-standard mediator because of the larger event rate with respect to atmospheric neutrinos. In particular, $1$~yr exposure to solar neutrinos could exclude a region of the parameter space similar to the one excluded by the observation of neutrinos from a galactic supernova burst. However, DARWIN will be more competitive on the solar neutrino bounds, given its better performance in the energy range of interest; while the supernova neutrino bounds will be driven by RES-NOVA-3 because of the higher energies of supernova neutrinos that are responsible for a higher event rate in lead.

Our results indicate that detectors employing lead and xenon targets are characterized by comparable sensitivity to non-standard interactions, given akin volumes. 
It is, however, important to notice that while the detection of solar and atmospheric neutrinos is connected to overcoming detector backgrounds, the observation of supernova neutrinos is essentially background free, thanks to the transient nature of the signal and the high expected event rate, at the price of being a rare event.
We have also tested the sensitivity of XENON1T to non-standard coherent scattering between neutrinos and nuclei. XENON1T performs better than  LUX~\cite{Akerib:2015rjg,Cerdeno:2016sfi}, due to its bigger effective volume, but the bounds obtained by COHERENT~\cite{AristizabalSierra:2019ykk} are more stringent. 

{
We have also demonstrated for the cases of a $U(1)_\mathrm{B-L}$ vector boson (Fig.~\ref{fig:BL}) and non-standard scalar coupling to neutrinos and quarks (Fig.~\ref{fig:scalar}) that DARWIN and RES-NOVA will be competitive with limits derived by relying on other arguments. In the near future new exciting possibilities of constraining non-standard mediators might come from the experiments focused on reactor neutrinos such as CONUS~\cite{Hakenmuller:2019ecb}, $\nu$-cleus~\cite{Strauss:2017cuu}, CONNIE~\cite{Aguilar-Arevalo:2019jlr}, MINER~\cite{Agnolet:2016zir}, and RED~\cite{Akimov:2012aya}.
In particular, CONUS is currently taking the data and its projected bounds in the new mediator parameter space may be competitive to the ones derived in this work~\cite{Farzan:2018gtr,Denton:2018xmq}. 
}

In conclusion, the possibility of detecting astrophysical neutrinos with coherent neutrino-nucleus scattering detectors opens a new window to explore New Physics. Excitingly, it also promises to place the most competitive bounds on non-standard neutrino-nucleus coherent scattering.

\acknowledgments

We are grateful to  Patrick Decowski, Rafael Lang, and Shashank Shalgar for helpful discussions and Diego Aristizabal Sierra for feedback on the manuscript. We also thank Robert Bollig, Thomas Ertl, and Thomas Janka for granting access to the data of the supernova models adopted in this work.
This work was supported by the Villum Foundation (Project No.~13164), the Carlsberg Foundation (CF18-0183), the Knud H\o jgaard Foundation, the Deutsche Forschungsgemeinschaft through Sonderforschungbereich SFB 1258 ``Neutrinos and Dark Matter in Astro- and Particle Physics'' (NDM).

\appendix

\section{Impact of the uncertainty in the supernova model, mass and distance}
\label{Appendix_B}

\begin{figure*}[t]
    \centering
    \includegraphics[scale=0.45]{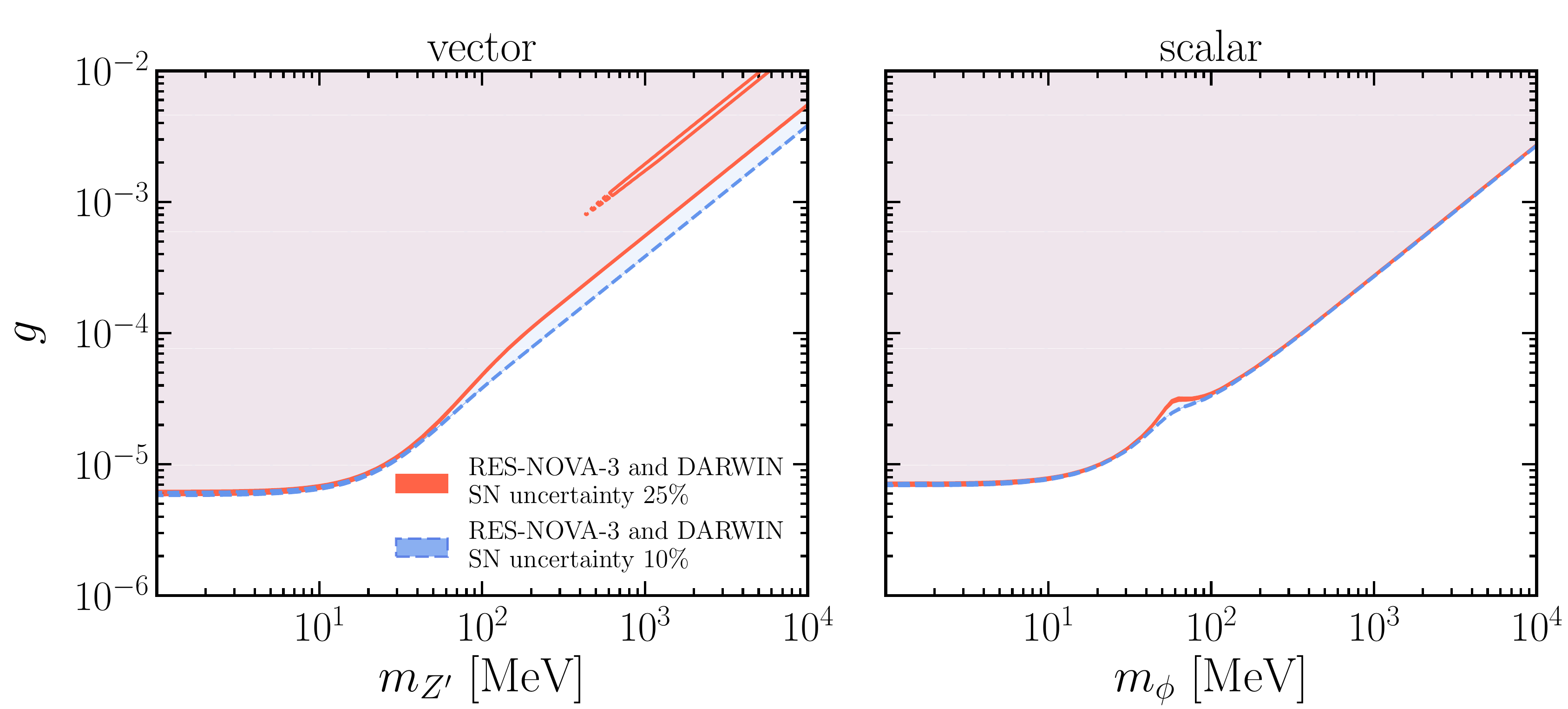}
    \caption{Projected 90$\%$~CL sensitivity bounds on the mass and coupling of the new vector (on the left) and scalar (on the right) mediators for a SN burst at  $10$~kpc from Earth detected in RES-NOVA-3 and DARWIN. The plotted bounds are marginalized over CC- and failed SNe. The solid (dashed) lines represent the bounds for the $1\sigma$ normalization uncertainty equal to $25\%$ ($10\%$). The $10\%$ uncertainty significantly helps  only for the case of  high mass vector mediator.}
    \label{fig:Contour_2D_SN_detector_combined}
\end{figure*}

To test the impact of the uncertainty in resolving the SN distance, and the possible uncertainty in the SN neutrino signal (expressed as an uncertainty on the signal normalization), we compute the projected $90\%$~CL exclusion regions for a SN at $10$~kpc from  Earth, assuming $10\%$ and $25\%$ uncertainty on the flux normalization. This normalization band includes  uncertainties in determining the SN distance~\cite{Kachelriess:2004ds,Adams:2013ana}, differences in the neutrino properties from various one-dimensional hydrodynamical SN simulations~\cite{OConnor:2018sti}, and uncertainties on the dependence of the neutrino emission properties from the SN mass~\cite{Seadrow:2018ftp}.

Figure~\ref{fig:Contour_2D_SN_detector_combined} shows the calculated bounds. The solid regions show results obtained by adopting the SN neutrino signal with a normalization uncertainty of $25\%$ and the dashed ones refer to the $10\%$ normalization uncertainty. Additionally, in order to take into account differences between the CC-SN and failed SN models (different shapes of the time integrated SN flux), the bounds are marginalized over the CC-SN and failed SN models. The left panel of Fig.~\ref{fig:Contour_2D_SN_detector_combined}  illustrates the expected bounds for the vector mediator case; the higher normalization uncertainty only affects  the high mass part of the contour plot. This is explained by the absence of event rate features for high masses of the vector mediator and the simple $(g^4 / m^2_{Z^\prime})$ or $(g^4 / m^4_{Z^\prime})$ effective coupling, which affect  the normalization of the resulting recoil rate. For the scalar mediator case (plot on the right of Fig.~\ref{fig:Contour_2D_SN_detector_combined}), because of the different kinetic term between  the scalar cross section and the vector one, the change of the normalization uncertainty does not affect the projected constraints significantly, see also Appendix~\ref{Appendix_A}.

\section{Dependence of the supernova neutrino rate on the mediator mass}
\label{Appendix_A}

To better understand the shape of the exclusion bounds for the vector and scalar mediators in Figs.~\ref{fig:2D_SN_BMS_10kpc}--\ref{fig:2D_SUN_ATM}, we here focus on the spectral features of the predicted recoil rates in three different mass regimes. 
\begin{figure*}[t]
    \centering
    \includegraphics[scale=0.53]{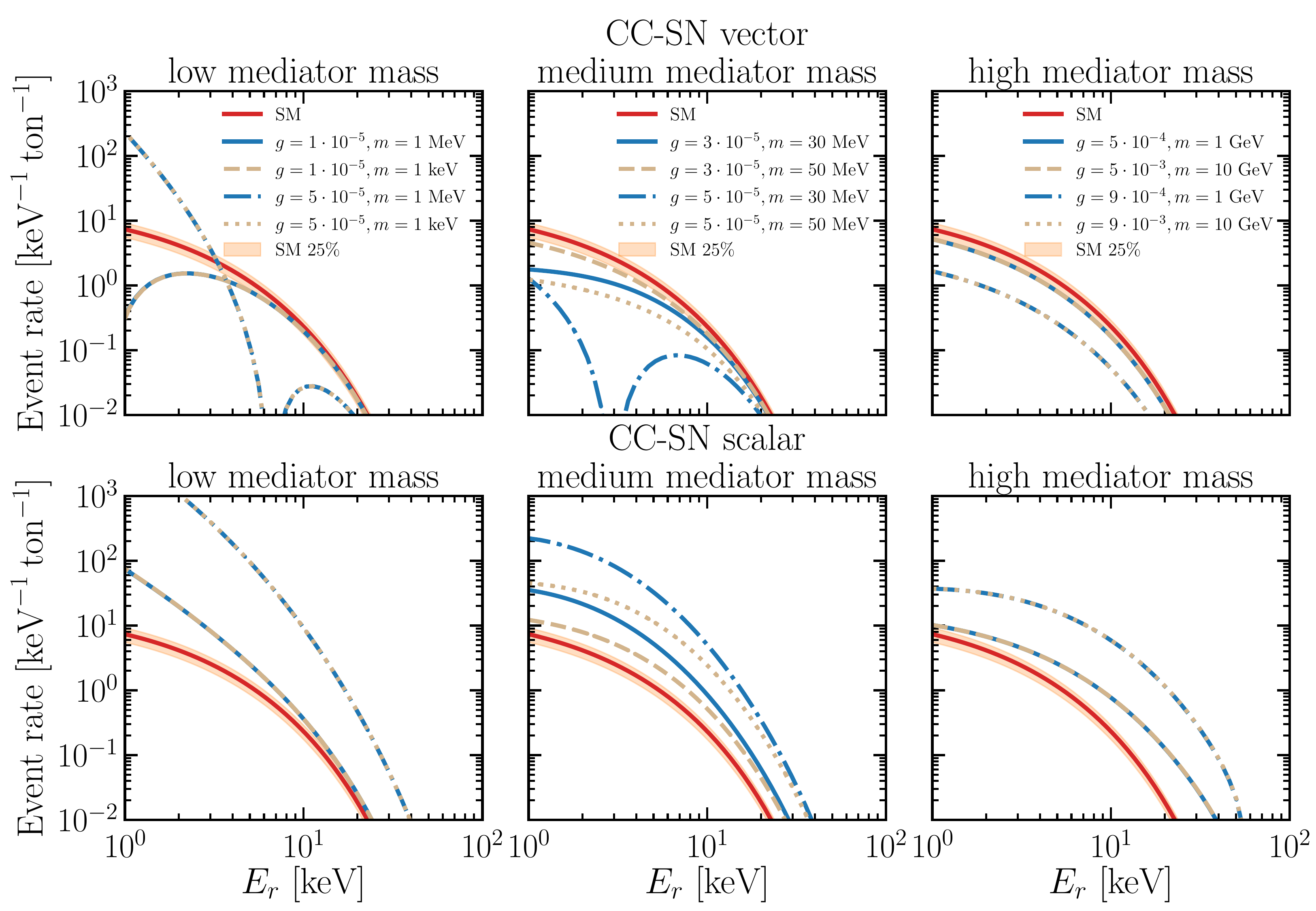}
    \caption{Expected event rate as a function of the nuclear recoil energy induced by the neutrino signal from a galactic supernova (CC-SN) at $10$~kpc per 
    $1$~ton of Xe detector for vector and scalar mediators are shown in blue and beige, in the top and bottom panels, respectively. The standard event rate (solid red lines) and its $25\%$ normalization uncertainty (orange band) is plotted for reference. The event rate in the case of low, medium and high mediator masses is plotted in the left, middle and right panels respectively. In the case of a  new scalar mediator, the rate  always results in higher recoil rates than in the standard scenario; this is not  the case for the vector mediator scenario.} 
    \label{fig:App}
\end{figure*}
Figure~\ref{fig:App} illustrates the event rates for a galactic supernova (CC-NS) at  $10$~kpc from Earth in a Xe based detector. The rates for the vector (scalar) mediator are in the top (bottom) panels. In each plot, the solid blue lines represent the event rate for the standard cross section and the orange band represents  its $25\%$ normalization uncertainty. The event rates in the presence of the new mediator are plotted as blue and beige lines for the same mediator mass and coupling, respectively. From Fig.~\ref{fig:App}, we  see that  the event rates for the scalar mediator are always larger than the standard recoil rate.

In the low mediator mass limit ($2 E_r m_T \gg m_{\{Z^\prime, \phi\}}^2$, left panels of Fig.~\ref{fig:App}) the event rates are insensitive to the mass of the  mediator independently of the mediator type. The recoil spectra for the intermediate mediator mass region ($2 E_r m_T \sim m_{\{Z^\prime, \phi\}}^2$, middle panels in Fig.~\ref{fig:App}) are sensitive to both  mass and coupling of the  mediator. 
In the low and medium mass regime (left and middle panels of Fig.~\ref{fig:App}), the event rates for the new vector mediator experience characteristic dips in the recoil spectrum due to the presence of the interference term between the standard and non-standard terms in Eq.~\ref{eq:sigma_total_vector}. The exact location of the dip depends on the new mediator coupling~\cite{AristizabalSierra:2019ykk}:
\begin{equation}
\label{eq:dip_location}
E_r = \frac{2g_{\nu, Z^\prime} Q_w^\prime - \sqrt{2} G_F m_{Z^\prime} Q_w}{2\sqrt{2} G_F Q_w m_T} \ .
\end{equation}

In the high mediator mass limit ($2 E_r m_T \ll m_{\{Z^\prime, \phi\}}^2$, right panels in Fig.~\ref{fig:App}), the non-standard couplings become effective couplings. For the vector mediator, the effective coupling is $\sim g^4 / m^2_{Z^\prime}$ when the interference term dominates and $\sim g^4 / m^4_{Z^\prime}$ when the non-standard term dominates. This effectively means that the total cross section (Eq.~\ref{eq:sigma_total_vector}) has the same energy dependence as the standard one, but  another normalization. For the scalar mediator case, since there is no interference term, the non-standard part of the total cross section (Eq.~\ref{eq:sigma_total_scalar}) will scale as $\sim g^4/m^4_{\phi}$. However, due to the different kinematic term with respect to the standard term, the new total cross section cannot be simply obtained by varying the cross section normalization.

\bibliography{NSI_SN.bib}
\end{document}